\documentclass[11pt]{article}
\usepackage[bookmarks=true,bookmarksnumbered=true,setpagesize=false]{hyperref}
\usepackage{a4wide}
\usepackage{amsmath,amssymb,amsfonts}
\usepackage{bbm}

\usepackage{comment}

\usepackage{enumerate}
\usepackage{url}
\usepackage{graphicx}

\usepackage{color}

\definecolor{darkblue}{rgb}{0,0,0}
\usepackage{braket}

\DeclareMathOperator{\diag}{diag}

\DeclareMathOperator{\PLC}{PLC}
\DeclareMathOperator{\FLC}{FLC}
\DeclareMathOperator{\FBC}{FBC}

\DeclareMathOperator{\arccosh}{arccosh}

\newcommand{\df}{\tx{d}}

\newcommand{\ov}{\over}

\newcommand{\bs}[1]{\boldsymbol{#1}}
\newcommand{\bh}[1]{\boldsymbol{\hat{#1}}}

\newcommand{\h}[1]{\hat{#1}}

\newcommand{\al}[1]{\begin{align}#1\end{align}}
\newcommand{\als}[1]{\begin{align*}#1\end{align*}}
\newcommand{\ol}{\overline}

\newcommand{\ab}[1]{\left|#1\right|}
\newcommand{\paren}[1]{\left(#1\right)}
\newcommand{\sqbr}[1]{\left[#1\right]}
\newcommand{\fn}[1]{\!\paren{#1}} 

\newcommand{\ttt}{\texttt{t}}

\newcommand{\bb}{\begin{bmatrix}}
\newcommand{\eb}{\end{bmatrix}}
\newcommand{\bmat}[1]{\begin{bmatrix}#1\end{bmatrix}}
\newcommand{\nn}{\nonumber\\}

\newcommand{\I}{\tx{I}}
\newcommand{\T}{\tx{T}}

\newcommand{\iq}{\mathbbm{i}}

\newcommand{\lip}[1]{\left[#1\right]}
\newcommand{\pr}{\prime}
\newcommand{\mc}{\mathcal}
\renewcommand{\P}{\text{P}}

\newcommand{\xP}{x_\P}
\newcommand{\xO}{x_\tx{O}}
\newcommand{\tauP}{\tau_\P}
\newcommand{\uP}{u_\P}
\newcommand{\uO}{u_\tx{O}}
\newcommand{\aP}{a_\P}
\newcommand{\AP}{A_\P}

\newcommand{\tx}{\text}

\newcommand{\Or}[1]{\mathcal O\!\left(#1\right)}

\newcommand{\fv}{\overrightarrow} 

\begin{document}
\title{Relativity for games
\bigskip\\ }
\author{
	Daiju Nakayama\thanks{E-mail: {\tt 42.daiju@gmail.com}, {\tt nakayama@cp.cmc.osaka-u.ac.jp}}
	\ and 
	Kin-ya Oda\thanks{E-mail: {\tt odakin@gmail.com}, {\tt odakin@phys.sci.osaka-u.ac.jp}}\bigskip\\
	\normalsize\it $^*$ e-Seikatsu Co., Ltd., 5-2-32, Minami-Azabu, Minato, Tokyo 106-0047, Japan\\
	\normalsize\it $^*$ Cybermedia Center, Osaka University, Osaka 560-0043, Japan\\
	\normalsize\it $^\dagger$ Department of Physics, Osaka University, Osaka 560-0043, Japan
	\bigskip\\
	}

\maketitle
\begin{abstract}\noindent
We present how to implement special relativity in computer games.
The resultant relativistic world shows the time dilation and Lorentz contraction exactly, not only for the player but also for all the nonplayer characters, who obey the correct relativistic equation of motion according to their own accelerations.
Causality is explicitly maintained in our formulation by use of the covariant velocities, proper times, worldlines, and light cones.
Faraway relativistic scenes can be accurately projected onto the skydome.
We show how to approximate a rigid body consisting of polygons, which is ubiquitous in computer games but itself is not a relativistically invariant object.
We also give a simple idea to mimic the Doppler effect within the RGB color scheme.
\end{abstract}

\newpage

\tableofcontents
\newpage

\section{Introduction}
The world is written in the language of quantum mechanics and relativity.
They together provide the basis for the Standard Model of particle physics and for standard cosmology; see, e.g., Refs.~\cite{RPP2014} and \cite{NASA} for reviews, respectively.
For large length scales,\footnote{
To be more precise, this should be rephrased to ``large length scales compared to the inverse of typical energy and momentum scales of the system in natural units.''
}
classical (i.e.\ nonquantum) mechanics works well.
If all the objects have much lower speeds than the speed of light, nonrelativistic mechanics suffices.\footnote{
Here and hereafter, we neglect the general-relativistic effects: When we say ``relativity'', it refers to special relativity.
(In any case, human beings have not yet managed to reconcile quantum mechanics and general relativity; see e.g.\ Refs.~\cite{superstring,Smolin:2004sx,Ambjorn:2012jv}.)
}
When both conditions are met, the ordinary Newtonian mechanics becomes a good effective description of nature.

Quantum mechanics is hard to visualize, as it is beyond our ordinary perception.
On the other hand, there is no problem in visualizing relativistic effects, which can be significant even within the reach of classical (nonquantum) mechanics.
The detection process of light by our eyes does not differ whether the light is emitted from the relativistic world or not.
Therefore, there remains room to visualize a classical but relativistic world in computer games if we manage to formulate the implementation of a relativistic world.
This is the goal of this work.

Up to now, virtually all games have been based on Newtonian mechanics, with only a few exceptions.
In Refs.~\cite{VelRap} and \cite{ASSL}, a player can traverse a relativistic world with three and four spacetime dimensions, respectively, without any reaction from the nonplayer characters (NPCs).
These two are the only games that have so far been found by the authors.

One may also find a good review on the visualization of relativistic effects in computers in Ref.~\cite{Wthesis}. Further developments after Ref.~\cite{Wthesis} can be found in Refs.~\cite{China2001,Benger2002,Deissler,njp08,Special_Relativistic_Visualization_by_Local_Ray_Tracing,1f0fb0f06e}.
In these works, what is implemented are the relativistic effects for a moving observer in a world that does not evolve in time in itself. In other words, one cannot affect the world in this kind of implementation.

\begin{figure}[tp]
\begin{center}
\includegraphics[width=0.49\textwidth]{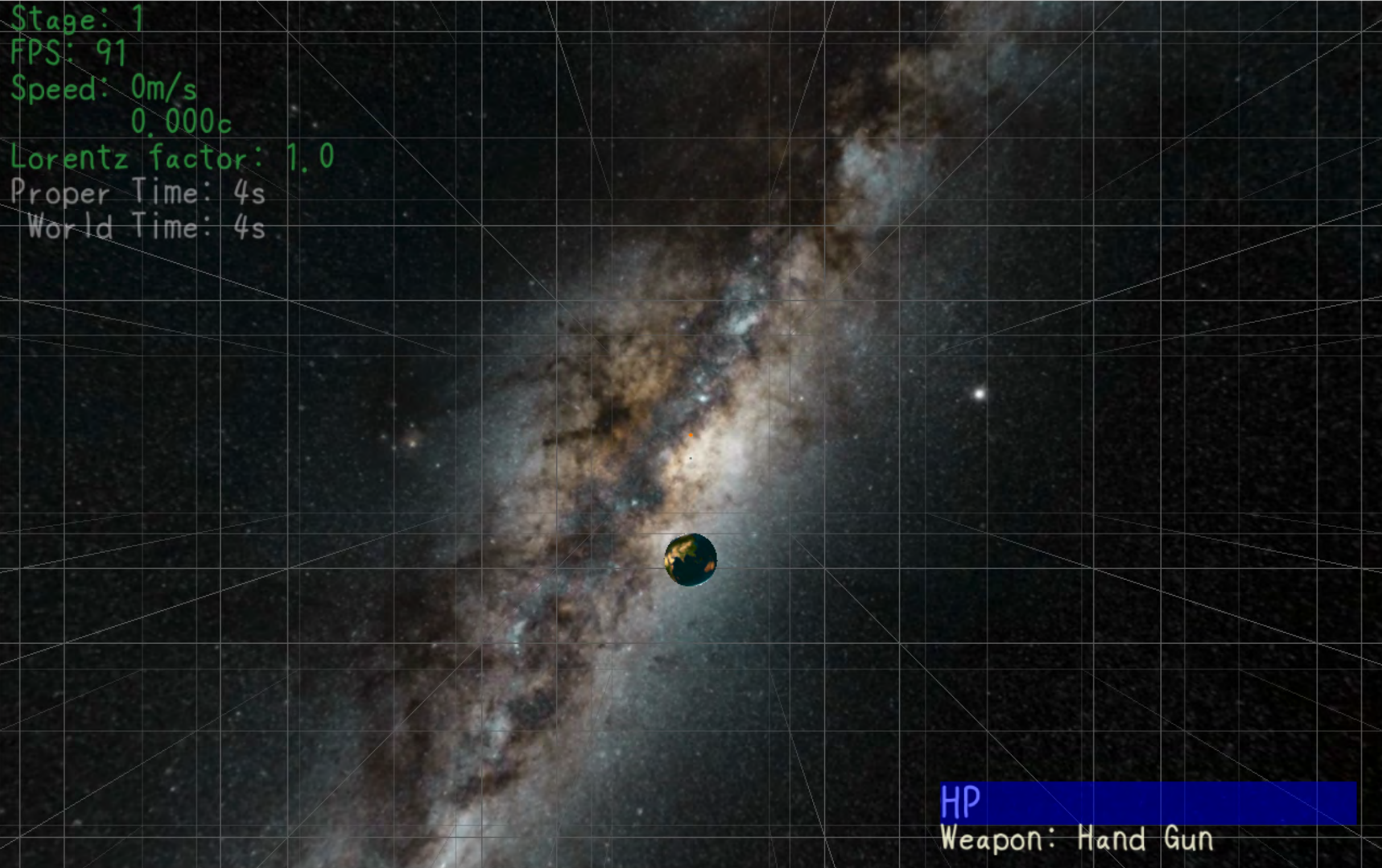}
\includegraphics[width=0.49\textwidth]{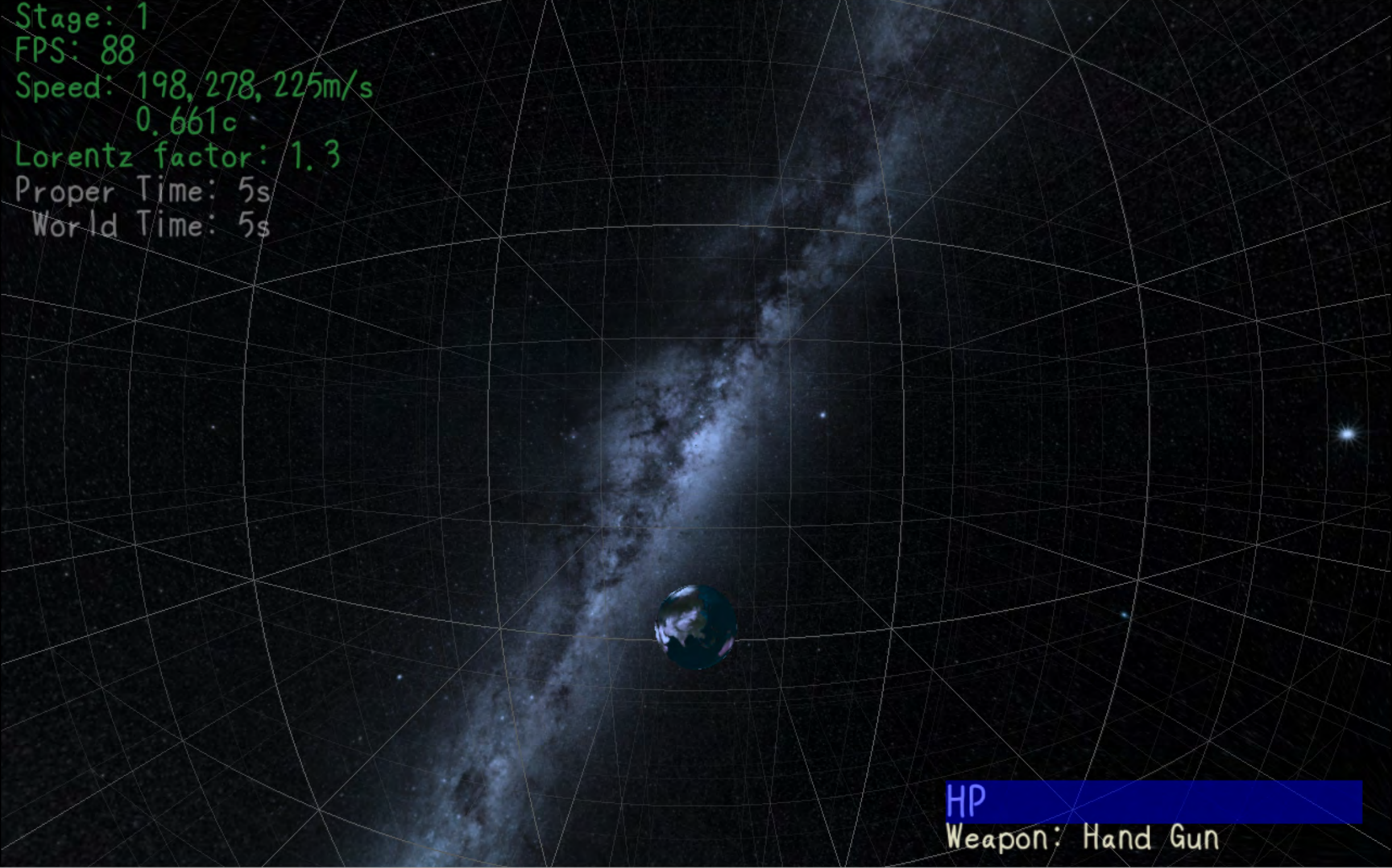}\\
\includegraphics[width=0.49\textwidth]{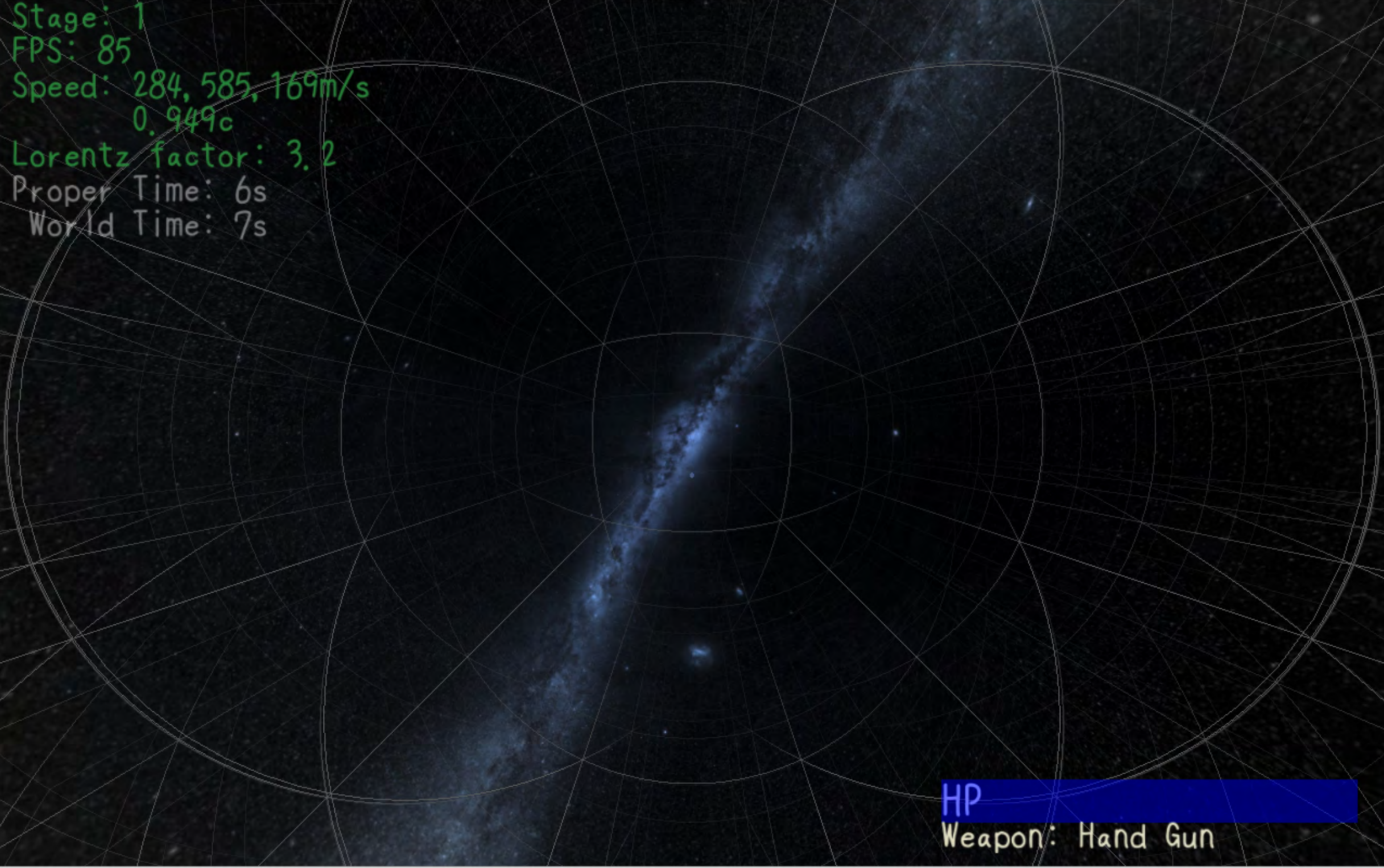}
\includegraphics[width=0.49\textwidth]{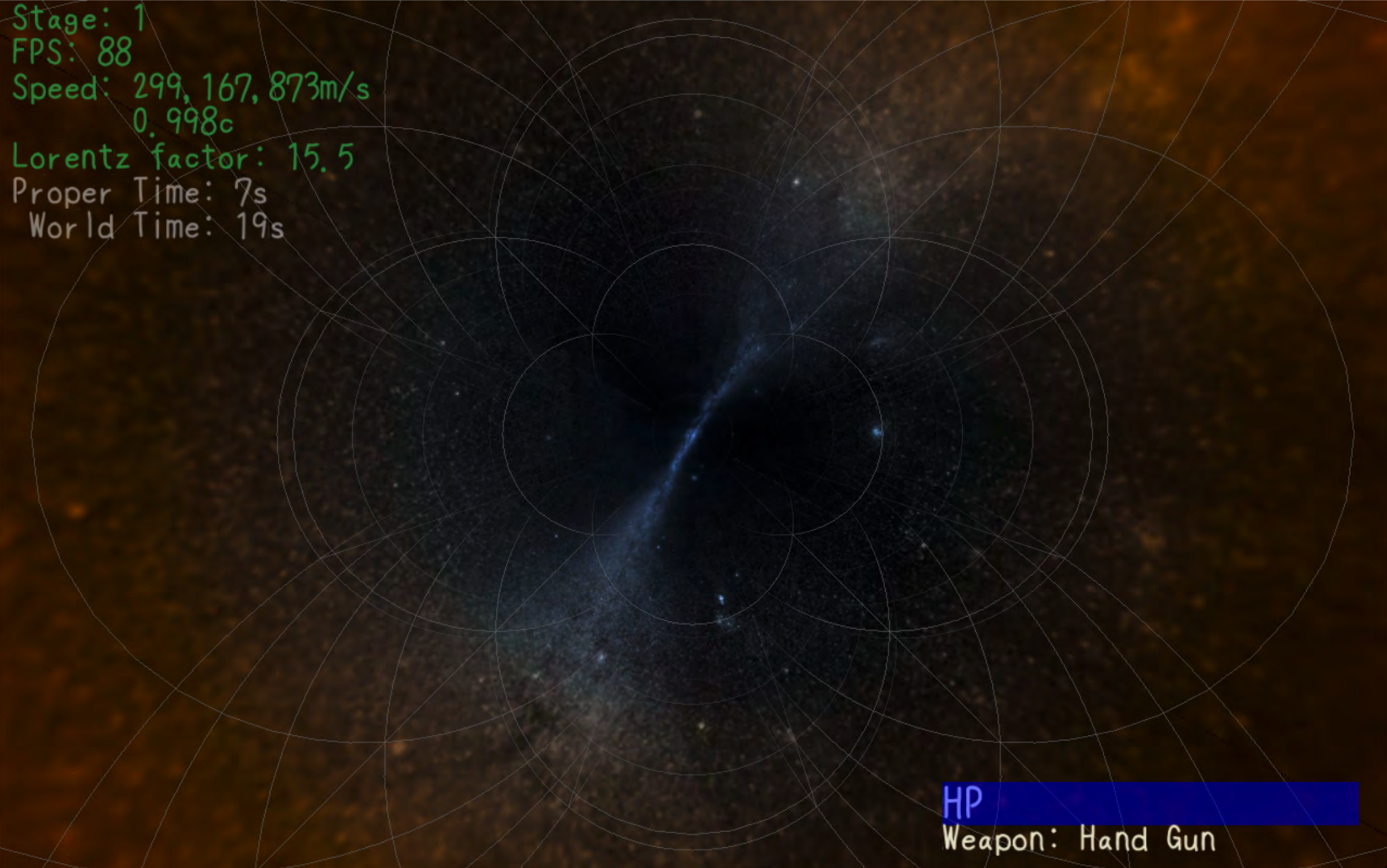}
\caption{\label{acceleration figure}
Game screens from the sample game in Ref.~\cite{sogebu}.
The player accelerates towards the center of the Milky Way Galaxy in the order of upper left, upper right, lower left, and lower right.
The background image of the Galaxy is Lorentz-transformed and drawn according to Sec.~\ref{skydome}.
(The image files for the Galaxy and Earth are taken from Refs.~\cite{milkyway} and \cite{earth}, respectively.)
The size of the Lorentz-noncovariant velocity $\bs v$ in Eq.~\eqref{noncovariant velocity} is 0 (upper left), $0.661\,c$ (upper right), $0.949\,c$ (lower left), and $0.998\,c$ (lower right), where $c$ is the speed of light.
The 3D world is gridded by the white lines, which are Cartesian coordinates in the reference frame introduced in Sec.~\ref{implementation}.
One can see that the coordinates appear bent as the player's velocity becomes sizable.
The Doppler effect is visible such that the color of the Galactic Center is blue-shifted as the player accelerates. On the other hand, one can see that the backward scene, which is squeezed into the player's sight due to the so-called relativistic aberration, is red-shifted on the lower-right panel.
Note that the implemented Doppler effect does not represent the real physical shift of the spectrum but is a mimicked one; see Sec.~\ref{Doppler section}.
We see that, as the player accelerates, the player's ``proper time'', corresponding to $\tau$ in Eq.~\eqref{proper time}, becomes delayed compared to the ``world time'' of the background at rest.
(``FPS'' in the second line on the screen denotes ``frames per second'' rather than ``first-person shooter''.)
}
\end{center}
\end{figure}

\begin{figure}[tp]
\begin{center}
\includegraphics[width=0.49\textwidth]{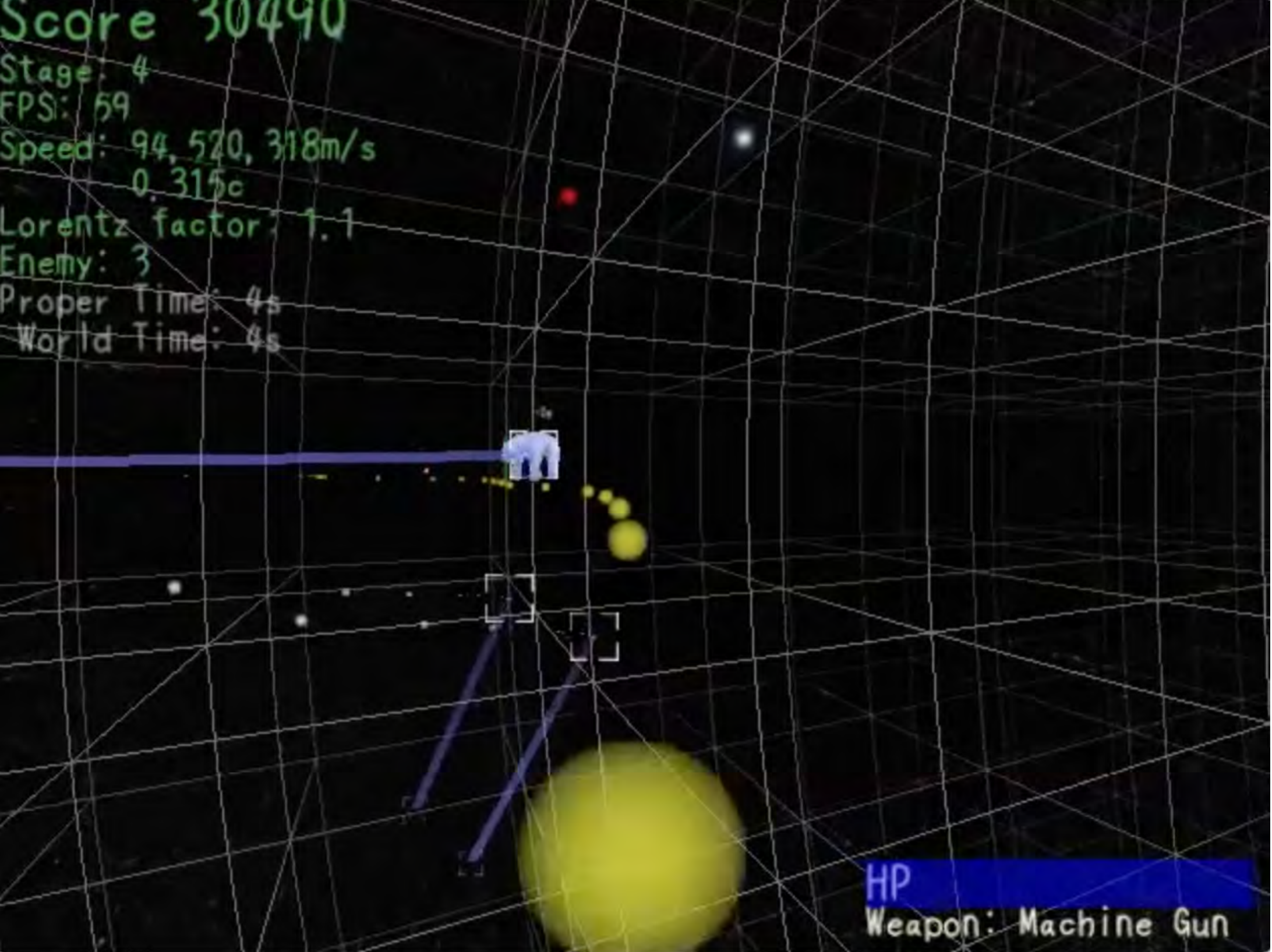}
\includegraphics[width=0.49\textwidth]{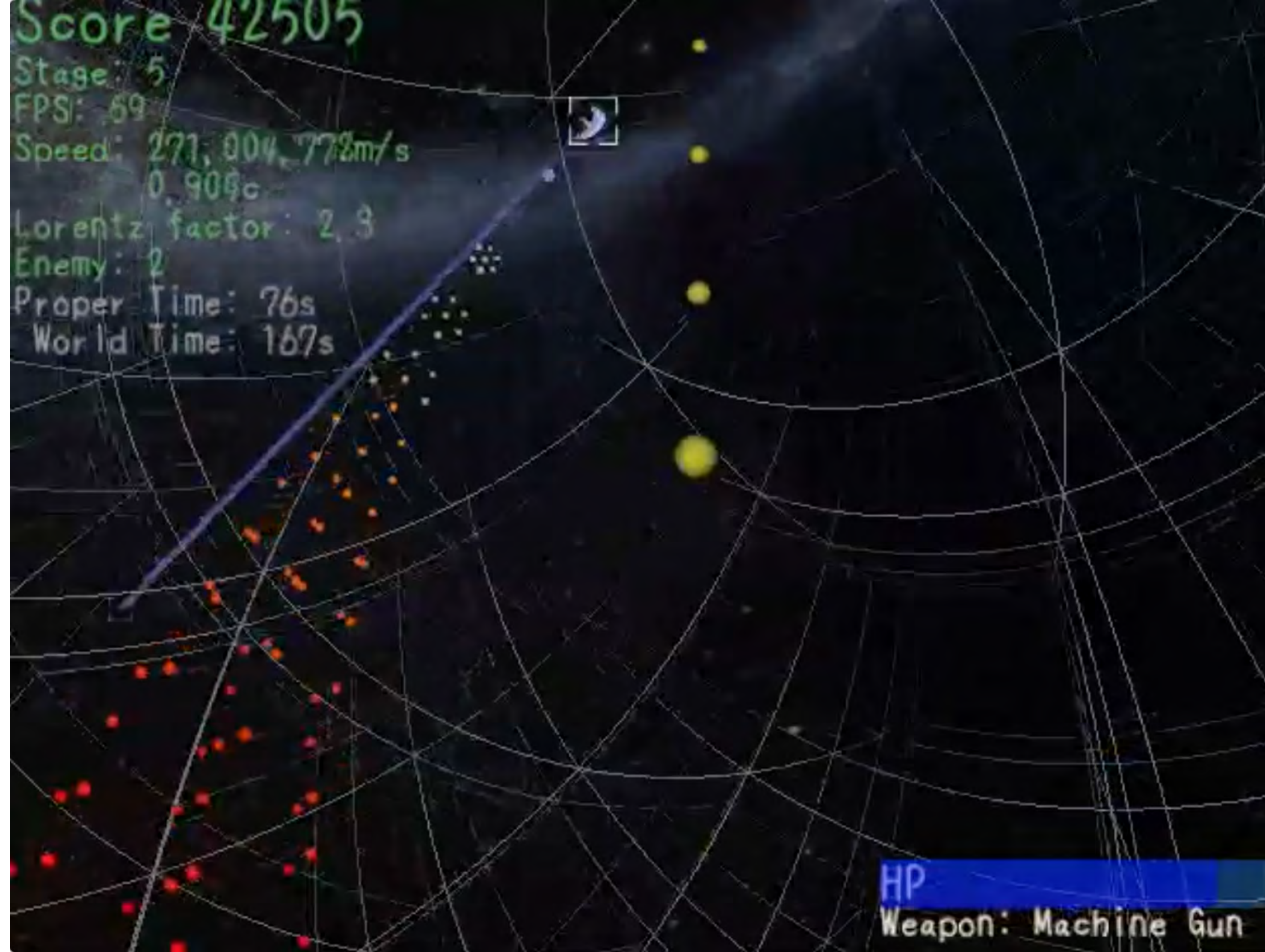}
\caption{\label{shooting figure}
Game screens from the sample game in Ref.~\cite{sogebu}.
The player is shooting the yellow bullets.
Both the enemy and bullet are assumed to emit light of given colors by themselves, rather than reflecting light from some source such as the Sun.
The player's velocity is faster on the right so that the Cartesian coordinates, explained in the caption of Fig.~\ref{acceleration figure}, appear more bent.
On the right, the enemy is shooting bullets towards the player according to the aiming algorithm in Sec.~\ref{sec:aiming}; we can also see the red shift on the enemy's bullets that come in the player's sight due to the relativistic aberration. We note again that the implemented Doppler effect does not represent the real physical shift of the spectrum but is a mimicked one; see Sec.~\ref{Doppler section}. The time evolution of the player and enemy is determined as described in Sec.~\ref{time evolution}: The enemy determines its move according to the information on its own past light cone as in Sec.~\ref{formulation}, and estimates others' move projected on its own future light cone as in Ref.~\ref{sec:aiming}. The larger and smaller square cursors, which are connected by the straight blue lines, indicate the enemy's position on the player's past light cone in Eq.~\eqref{new location} and the enemy's projected direction on the future light cone in Eq.~\eqref{future projection in rest frame}, respectively.
}
\end{center}
\end{figure}

In this paper, we formulate how to implement time evolution in a special-relativistic world,
namely, interactivity between the NPCs, the player, and all the other objects.
In particular, one will see that causality is maintained by heavy use of past light cones and worldlines in four spacetime dimensions.
One may find concrete implementation of all the ideas in this paper in a primitive first-person shooter (FPS) game in Ref.~\cite{sogebu}.
Sample game screens are shown in Fig.~\ref{acceleration figure}, in which the player is accelerating in a fixed background without affecting the world.
In Fig.~\ref{shooting figure}, the player is fighting against enemies interactively.

This paper is organized as follows.
In Sec.~\ref{special relativity}, we review special relativity and spell out our notations.
In Sec.~\ref{implementation}, we show how to draw the fully special-relativistic world in computer games. Each observed object is drawn at the intersection between the observer's past light cone and the object's worldline.
In Sec.~\ref{time evolution}, we present how to evolve the relativistic world, with an application to an FPS in mind. This is the first time that interactions between participants (namely, the player and the NPCs) are consistently taken into account without contradicting causality. Past and future light cones are heavily used for this.
In Sec.~\ref{miscellaneous}, we give various ideas useful for a concrete implementation.
In Sec.~\ref{summary}, we summarize our result.
In Appendix~\ref{energy momentum}, we briefly summarize how the energy and momentum form a spacetime vector in relativity.
In Appendix~\ref{Lorentz contraction and dilation}, we show how the Lorentz contraction and \emph{dilation} occur when a measuring rod is seen by an observer.
In Appendix~\ref{co-moving evolution}, we present another formulation for the time evolution of the objects on the player's past light cone. This co-moving evolution is more suitable for analytic treatment, and some of the results are shown too.
In Appendix~\ref{rocket propulsion}, as a concrete example of the relativistic acceleration, we show how the rocket propulsion can be formulated.
In Appendix~\ref{quaternion}, we review the use of quaternions in order to handle the rotation in three spatial dimensions.

We have tried to make this paper more pedagogical and self-contained than ordinary physics papers, even for some elementary mathematics, since the subject is more interdisciplinary and we expect more computer-oriented readers.

\section{Special relativity}\label{special relativity}
We spell out our basic notations to describe a special-relativistic world.
\subsection{Coordinates, vectors, worldlines}
Usually computer games have $d=2$ or 3 spatial dimensions.
We write a position vector in the $d$-dimensional Cartesian (orthonormal) coordinate system as\footnote{
In terms of the ordinary $(x,y,z)$ notation, which is not used in this paper unless otherwise stated, they are written as $\paren{x^1,\,x^2,\,x^3}=\paren{x,y,z}$.
}
\al{
\bs x
	=	\paren{x^1,\,\dots,\,x^d}
	=	\bmat{x^1\\ \vdots\\ x^d},
		\label{d-vector}
}
where the $d$ vector is written in the \emph{component} and \emph{matrix} notations in the second and third expressions, and we use curly and square brackets for them, respectively, unless otherwise stated.
We generally write a $d$ vector with a bold-style letter such as $\bs x$.
In the matrix notation, the transpose is given by
\al{
\bs x^\T
	&=	\bmat{x^1&\cdots&x^d}.
}
The location of each particle in the world is represented by its position vector $\bs x$ at a given time~$t$.
In practice, in computer games, the ``particle'' may stand for a vertex in a polygon.

The time dimension $t$ is written as the 0th coordinate\footnote{
Throughout this paper, ``$:=$'' indicates that the left-hand side is defined by the right-hand side, and vice versa for ``$=:$''.
}
\al{
x^0:=ct,
	\label{0th coordinate}
}
where $c=299\,792\,458\,\tx{m}/\tx{s}$ is the \emph{speed of light}, which is a physical constant in special relativity.\footnote{\label{different speed of light}
In computer games, one may instead choose any value of $c$ to simulate a world with a slower speed of light, as in Ref.~\cite{ASSL}.
}
We can always take the \emph{natural units} $c=1$, which we employ in this paper unless otherwise stated.
In the natural units, time and length have the same \emph{dimension}: $1\,\tx{s}=299\,792\,458\,\tx{m}$.
When necessary, we can always come back from the natural units by recovering $c$ by \emph{dimensional analysis}.

A unit length in computer programming is quite arbitrary, and we call it a ``pixel'' here just for concreteness.
We may identify a pixel with an arbitrary physical length $\ell$; namely, $\ell$ is given in the units of $\tx{m}/\tx{pixel}$. Then the speed of light in the program is
\al{
c={299\,792\,458\,\tx{m}/\tx{s}\ov \ell},
	\label{c in pixel}
}
which has units of $\tx{pixel}/\tx{s}$.\footnote{
One may instead take whatever value of the speed of light in the unit of $\tx{pixel}/\tx{s}$, regardless of the choice of $\ell$, if one simulates a different world with a different speed of light, as discussed in footnote~\ref{different speed of light}.
}
In practice, it is most convenient to store all the time information in pixels too, by using the speed of light~\eqref{c in pixel} in Eq.~\eqref{0th coordinate}.
Then the velocity and acceleration, which will be presented in Eqs.~\eqref{covariant velocity given} and \eqref{acceleration given}, are given without unit and with units of $\tx{pixel}^{-1}$, respectively.\footnote{
When one only uses the acceleration to change the velocity, as in Secs.~\ref{player's time evolution} and \ref{others' time evolution}, it would be more practical to give the proper-time difference $\Delta\tau$ in units of seconds. Then the time $x^0$ (as well as the position $\fv x$), the velocity $\fv u$, and the acceleration $\fv a$ would be given in units of pixels, dimensionless, and in the units of $\tx{s}^{-1}$, respectively. In that case, the time evolution in the nonnatural units becomes schematically
\als{
\tau	&\longmapsto	\tau+\Delta\tau,\\
\fv x	&\longmapsto	\fv x+\fv u\,c\,\Delta\tau,\\
\fv u	&\longmapsto	\fv u+\fv a\Delta\tau,
}
where $c$ is given in units of $\tx{pixel}/\tx{s}$ as in Eq.~\eqref{c in pixel}.
}

The spacetime coordinates are collectively written as a $D$-dimensional vector with $D=d+1$:
\al{
\fv x
	&=	\paren{x^0,\,\bs x}
	=	\bmat{x^0\\ \bs x}\nn
	&=	\paren{x^0,\,x^1,\,\dots,\,x^d}
	=	\bmat{x^0\\ x^1\\ \vdots\\ x^d},
		\label{D-vector}
}
where we have again used the component and matrix notations in the second and third expressions in each line, respectively. We place the long arrow on top of any $D$ vector, such as $\fv x$, throughout this paper.
The transpose in the matrix notation is
\al{
\fv x^\T
	&=	\bmat{x^0&x^1&\cdots&x^d}
	=	\bmat{x^0&\bs x^\T}.
}

In ordinary Newtonian mechanics, the trajectory of each particle from a time $t_\tx{ini}$ to $t_\tx{fin}$ can be written as
\al{
\mc T	&=	\Set{\bs x\fn{t}|t_\tx{ini}\leq t\leq t_\tx{fin}}.
}
This trajectory can be viewed as a \emph{worldline} in $D$-dimensional spacetime:
\al{
W	&=	\Set{\fv x\fn{s}|s_\tx{ini}\leq s\leq s_\tx{fin}},
	\label{continuous worldline}
}
where $\fv x$ is the $D$ vector given in Eq.~\eqref{D-vector} and the real number $s$ parametrizes the worldline.\footnote{\label{affine footnote}
One takes $s$ to be an \emph{affine parameter} that is invariant under \emph{symmetries} that one wants to impose.
In the current Newtonian case, one may (but does not have to) identify the parameter $s$ as the time coordinate $t$ itself so that $s$ is defined  rotation-invariantly. Instead we may also identify $s$ as an arbitrary monotonically increasing function of $t$. This freedom is called the \emph{reparametrization invariance}.
}

\subsection{Newtonian mechanics}
In Newtonian mechanics, let us consider a particle at time $t$ located at $\bs x\fn{t}$, having a (nonrelativistic) velocity
\al{
\bs v\fn{t}
	&=	\paren{v^1\fn{t},\,\dots,\,v^d\fn{t}}
	=	\bmat{v^1\fn{t}\\ \vdots\\ v^d\fn{t}}.
}
The particle's position after an infinitesimal time $\Delta t$ is given by
\al{
\bs x\fn{t+\Delta t}
	&=	\bs x\fn{t}+\bs v\fn{t}\Delta t,
}
or, to be more explicit,
\al{
\bmat{x^1\fn{t+\Delta t}\\ \vdots\\ x^d\fn{t+\Delta t}}
	&=	\bmat{x^1\fn{t}\\ \vdots\\ x^d\fn{t}}
		+\bmat{v^1\fn{t}\\ \vdots\\ v^d\fn{t}}
			\Delta t\nn
	&=	\bmat{
			x^1\fn{t}+v^1\fn{t}\Delta t\\
			\vdots\\
			x^d\fn{t}+v^d\fn{t}\Delta t
			}.
}
The velocity $\bs v$ at each moment can be chosen quite arbitrarily in nonrelativistic computer games.
If we use Newtonian mechanics, $\bs v\fn{t+\Delta t}$ is determined from $\bs v\fn{t}$ for a given input of the (nonrelativistic) acceleration $\bs a_\tx{NR}\fn{t}$:
\al{
\bs v\fn{t+\Delta t}
	&=	\bs v\fn{t}+\bs a_\tx{NR}\fn{t}\Delta t,
		\nn
\bmat{v^1\fn{t+\Delta t}\\ \vdots\\ v^d\fn{t+\Delta t}}
	&=	\bmat{v^1\fn{t}\\ \vdots\\ v^d\fn{t}}
		+\bmat{a_\tx{NR}^1\fn{t}\\ \vdots\\ a_\tx{NR}^d\fn{t}}
			\Delta t,
}
where the acceleration at $t$ is determined by Newton's equation of motion:
\al{
m\,\bs a_\tx{NR}\fn{t}
	&=	\bs F\fn{t}.
		\label{Newtonian eom}
}
The explicit form of the force $\bs F$ depends on the dynamics that you choose.

In computer games, the trajectory and worldline become discrete sets of $d$ and $D$ vectors, respectively, due to discrete time steps:
\al{
\mc T	&=	\Set{\bs x\fn{t_1},\,\bs x\fn{t_2},\,\dots,\,\bs x\fn{t_N}},\\
W	&=	\Set{\fv x\fn{s_1},\,\fv x\fn{s_2},\,\dots,\,\fv x\fn{s_N}}.
	\label{Newtonian discrete worldline}
}

\subsection{Lorentz transformation}
In spatial $d$ dimensions, the ordinary inner product of a pair of $d$ vectors, say $\bs x=\paren{x^1,\dots,x^d}$ and $\bs y=\paren{ y^1,\dots, y^d}$, is given by
\al{
\bs x\cdot\bs{ y}
	&:=	\sum_{i=1}^dx^i y^i\nn
	&=	x^1y^1+\cdots+x^dy^d.
		\label{Euclidean product}
}
We may also write
\al{
\bs x\cdot\bs y
	=	\bs x^\T\,\bs y
	&=	\bmat{x^1&\cdots&x^d}\bmat{y^1\\ \vdots\\ y^d}
}
in the matrix notation. Note that $\bs x^\T\, \bs y=\paren{\bs x^\T\, \bs y}^\T=\bs y^\T\,\bs x$.

We define the \emph{Lorentzian inner product} of the $D$ vectors $\fv x=\paren{x^0,\,x^1,\,\dots,\,x^d}$ and $\fv{ y}=\paren{ y^{0},\, y^{1},\,\dots,\, y^{ d}}$ by\footnote{
A dot product of $D$ vectors always denotes the Lorentzian inner product~\eqref{lip}, and never the Euclidean one~\eqref{Euclidean product}.
}
\al{
\fv x\cdot\fv{ y}
	&:=	-x^0 y^{0}+\sum_{i=1}^dx^i y^{ i}.
		\label{lip}
}
Note that $\fv x\cdot\fv{ y}=\fv{ y}\cdot\fv x$.
We also use the notation
\al{
\lip{\fv x,\fv{ y}}
	&:=	\fv x\cdot\fv{ y},
}
interchangeably.

The Lorentzian inner product can be conveniently written in terms of the \emph{metric} $\eta$:
\al{
\fv x\cdot\fv{ y}
	&=	\fv x^\T\,\eta\,\fv{ y}
	=	\sum_{\mu,\nu=0}^dx^\mu\,\eta_{\mu\nu}\,y^\nu\nn
	&=	\bmat{x^0&x^1&\cdots&x^d}
		\bmat{
			\eta_{00}&\eta_{01}&\cdots&\eta_{0d}\\
			\eta_{10}&\eta_{11}&\cdots&\eta_{1d}\\
			\vdots   &\vdots   &\ddots&\vdots\\
			\eta_{d0}&\eta_{d1}&\cdots&\eta_{dd}
			}
		\bmat{ y^0\\  y^1\\ \vdots\\  y^d}\nn
	&=	\bmat{x^0&x^1&\cdots&x^d}
		\bmat{
			-1\\
			&1\\
			&&\ddots\\
			&&&1
			}
		\bmat{ y^0\\  y^1\\ \vdots\\  y^d},
		\label{lip in matrix}
}
where 
\al{
\eta_{\mu\nu}
	&:=	\begin{cases}
		-1	&	(\mu=\nu=0),\\
		1	&	(\mu=\nu=1,\dots,d),\\
		0	&	\tx{otherwise}.
		\end{cases}\label{metric}
}
Matrix-wise, we may also write $\eta=\diag\fn{-1,1,\dots,1}$, where ``$\diag$'' denotes a diagonal matrix. 

We also define the \emph{Lorentzian norm}:\footnote{
The extra square bracket is placed for a reader unfamiliar with relativistic notation:
Do not confuse the Lorentzian norm $\lip{\fv x}^2$ with the 2-component of $\fv x$ which is denoted by $x^2$.
}
\al{
\lip{\fv x}^2
	:=	\lip{\fv x,\,\fv x}
	=	\fv x\cdot\fv x
	&=	\fv x^\T\,\eta\,\fv x
	=	\sum_{\mu,\nu=0}^dx^\mu\,\eta_{\mu\nu}\,x^\nu\nn
	&=	-\paren{x^0}^2+\paren{x^1}^2+\cdots+\paren{x^d}^2\nn
	&=	-\paren{x^0}^2+\bs x^2,
		\label{Lorentzian norm}
}
where
\al{
\bs x^2:=\bs x\cdot\bs x=\sum_{i=1}^d\paren{x^i}^2
}
is the ordinary $d$-dimensional (squared) norm;
we may also write $\bs x^2=\bs x^\T\,\bs x$ in the matrix notation;
we also write
\al{
\ab{\bs x}
	&:=	\sqrt{\bs x^2}\geq0.
}
Note that the Lorentzian norm~\eqref{Lorentzian norm} can be negative. Any $D$ vector $\fv x$ is called \emph{spatial} (\emph{space-like}), \emph{null} (\emph{light-like}), and \emph{temporal} (\emph{time-like}) when $\lip{\fv x}^2$ is positive, zero, and negative, respectively:
\al{
\lip{\fv x}^2
	\begin{cases}
	>0	&	\text{spatial (space-like)},\\
	=0	&	\text{null (light-like)},\\
	<0	&	\text{temporal (time-like)}.
	\end{cases}
}
It is straightforward to show that the Lorentzian norm behaves in the same way as the ordinary square for a sum of any $D$ vectors:
\al{
\lip{\fv x+\fv y}^2
	&=	\lip{\fv x}^2+2\lip{\fv x,\fv y}+\lip{\fv y}^2.
}

Let us consider a linear coordinate transformation $\Lambda$:\footnote{\label{Poincare}
More generally, we impose the invariance under the \emph{Poincar\'e transformation}:
\als{
\fv x\to\Lambda\fv x+\fv b.
}
The addition of a constant vector $\fv b$ is the \emph{translation}, a constant shift of the coordinate origin, which is trivially realized and is not treated in this paper. A reader unfamiliar with the translation should only memorize the fact that any subtraction of two coordinates is \emph{invariant} under any translation: $\fv x-\fv y\to\fv{x'}-\fv{y'}=\paren{\fv x+\fv b}-\paren{\fv y+\fv b}=\fv x-\fv y$, and so is the covariant velocity~\eqref{covariant velocity given}. As the velocity is invariant under translation, the acceleration~\eqref{acceleration given}  becomes trivially invariant too.
}
\al{
\fv x	
	&\to	\fv{x'}=\Lambda \fv x,	\label{Lambda}\\
x^\mu
	&\to	x^{\pr\mu}
			=\sum_{\nu=0}^d\Lambda^\mu{}_\nu\,x^\nu\quad\tx{for $\mu=0,\dots,d$},\nn
\bmat{x^0\\ x^1\\ \vdots\\ x^d}
	&\to	\bmat{x^{\pr0}\\ x^{\pr1}\\ \vdots\\ x^{\pr d}}
			=
			\bmat{
				\Lambda^0{}_0&\Lambda^0{}_1&\cdots&\Lambda^0{}_d\\
				\Lambda^1{}_0&\Lambda^1{}_1&\cdots&\Lambda^1{}_d\\
				\vdots       &\vdots       &\ddots&\vdots\\
				\Lambda^d{}_0&\Lambda^d{}_1&\cdots&\Lambda^d{}_d
				}
			\bmat{x^0\\ x^1\\ \vdots\\ x^d},\nonumber
}
where we have also shown more explicit component expressions in the second and third lines; $\Lambda$ is a matrix,\footnote{
A reader unfamiliar with the distinction between the upper and lower indices does not have to be bothered by it. In this paper, we do not raise and lower the time index 0. We can freely raise and lower the spatial indices $1,\dots,d$ under the metric convention~\eqref{metric}. That is, $\Lambda^i{}_j=\Lambda_{ij}=\Lambda^{ij}$ for $i,j=1,\dots,d$.
}
\al{
\Lambda
	&=	\bmat{\Lambda^\mu{}_\nu}_{\mu,\nu=0,1,\dots,d}.
}
Note that all the $D$ vectors are transformed by Eq.~\eqref{Lambda} simultaneously: In particular $\fv x$ and $\fv{ y}$ are transformed by the same $\Lambda$.

The transformation~\eqref{Lambda} is called the \emph{Lorentz transformation} if it does not change the Lorentzian inner product~\eqref{lip}:
\al{
\fv{x'}\cdot\fv{ y'}
	=	\fv x\cdot\fv{ y},
}
which is satisfied when and only when\footnote{\label{rotation footnote}
The condition~\eqref{Lorentz condition} is analogous to the condition for the spatial rotation $R^\T R=\I$, which is deduced from the rotational invariance $\bs x\cdot\bs y\to \bs x'\cdot\bs y'=\bs x\cdot\bs y$ under the coordinate transformation $\bs x\to \bs x'=R\,\bs x$.
}
\al{
\Lambda^\T\,\eta\,\Lambda
	&=	\eta,		\label{Lorentz condition}
}
where\footnote{
In the component language, Eq.~\eqref{Lorentz condition} reads
\als{
\sum_{\mu',\nu'=0}^d\Lambda^{\mu'}{}_\mu\,\eta_{\mu'\nu'}\,\Lambda^{\nu'}{}_\nu
	&=	\eta_{\mu\nu},
}
for $\mu,\nu=0,\dots,d$.
}
\al{
\Lambda^\T
	&=		\bmat{
				\Lambda^0{}_0&\Lambda^1{}_0&\cdots&\Lambda^d{}_0\\
				\Lambda^0{}_1&\Lambda^1{}_1&\cdots&\Lambda^d{}_1\\
				\vdots       &\vdots       &\ddots&\vdots\\
				\Lambda^0{}_d&\Lambda^1{}_d&\cdots&\Lambda^d{}_d
				}.
}
This can be shown as follows:
\al{
\fv{x'}\cdot\fv{ y'}
	&=	\paren{\Lambda\fv x}\cdot\paren{\Lambda\fv{ y}}
	=	\paren{\Lambda\fv x}^\T\eta\,\paren{\Lambda\fv{ y}}
	=	\fv x^\T\paren{\Lambda^\T\,\eta\,\Lambda}\fv y,
}
where the matrix notation is used in the second step as in Eq.~\eqref{lip in matrix}.
That is, $\Lambda$ is a Lorentz transformation when and only when Eq.~\eqref{Lorentz condition} is satisfied.
It is important that any Lorentzian norm, say $\sqbr{\fv x}^2$, is then \emph{Lorentz-invariant}:
\al{
\sqbr{\fv x}^2	&\to	\sqbr{\fv x'}^2=\sqbr{\Lambda\fv x}^2=\sqbr{\fv x}^2.
}

Physically, we are looking for linear transformations that leave the speed of light invariant.
If the speed of light is invariant, then a spherical wavefront of light that is emitted from the same point should be transformed to a spherical one.
The wavefront of a spherical wave of light emitted from the origin at $t=0$ is represented by $\lip{\fv x}^2=0$.
We see that the Lorentz transformation indeed leaves it spherical:\footnote{
The \emph{dilatation} $\fv x\to b\fv x$, with $b\neq0$ being a constant, also leaves the condition $\lip{\fv x}^2=0$ unchanged. The invariance under dilations, called the \emph{scale invariance}, is violated by the \emph{conformal anomaly} in quantum field theories in general. Otherwise, the dilatation is excluded from the \emph{symmetry} of the system of our consideration under the following two assumptions~\cite{Einstein:1905ve}: First, a \emph{boost} along a direction with a given velocity $v$, combined with another boost along the same direction but with the negative velocity $-v$, becomes an identity transformation. Second, the length of a measuring rod perpendicular to the boost direction becomes identical to each other when its boost is with the velocity $v$ and $-v$.
}
$\lip{\fv{x'}}^2=\lip{\fv x}^2=0$.

In general, $\Lambda$ has $D\paren{D-1}/2=d\paren{d+1}/2$ degrees of freedom, and can be parametrized as
\al{
\Lambda
	&=	L\,\mc R,
}
where $L$ is a \emph{boost} matrix and
\al{
\mc R
	&=	\bmat{
			1&0&\cdots&0\\
			0&R_{11}&\cdots&R_{1d}\\
			\vdots&\vdots&\ddots&\vdots\\
			0&R_{d1}&\cdots&R_{dd}
			}
	=	\bmat{1&\bs 0^\T\\ \bs 0&R},
		\label{rotation matrix}
} 
in which $R$ is an ordinary rotation matrix in $d$ spatial dimensions and $\bs 0^\T=\bmat{0&\cdots&0}$. 
We note that $R$ obeys
\al{
R^\T R=\I,
}
where $\I$ is the $d$-dimensional identity matrix; see also footnote~\ref{rotation footnote}.
The $d$-dimensional rotation matrix $R$ has ${d\paren{d-1}\over2}$ degrees of freedom.
The remaining ${d\paren{d+1}\over2}-{d\paren{d-1}\over2}=d$ degrees of freedom in $\Lambda$ are in the boost matrix~$L$.
A concrete parametrization of $L$ will be given in Sec.~\ref{rest frame section} after we introduce more physics.

\subsection{Proper time and Lorentz-covariant velocity}\label{proper time and covariant velocity}
Special relativity is defined as a theory that is \emph{invariant} under Lorentz transformations.
Therefore, it is important to write the theory in terms of \emph{Lorentz-invariant} quantities.
A Lorentz-invariant quantity can be formed as a Lorentzian inner product~\eqref{lip in matrix} or a norm~\eqref{Lorentzian norm}. They are composed from $D$ vectors, which are \emph{Lorentz covariant}.
Therefore, it is important that things, such as the velocity and acceleration, are written as Lorentz-covariant $D$ vectors.

We start from a coordinate system $\Set{\fv x}$.
Suppose that an infinitesimal shift of the worldline parameter $s$ in Eq.~\eqref{continuous worldline} changes the position of the particle in $D$ spacetime dimensions:
\al{
\fv x\fn{ s}
	&\longmapsto	\fv x\fn{ s+\Delta s},\nn
\bmat{x^0\fn{ s}\\ x^1\fn{ s}\\ \vdots\\ x^d\fn{ s}}
	&\longmapsto	\bmat{x^0\fn{ s+\Delta s}\\ x^1\fn{ s+\Delta s}\\ \vdots\\ x^d\fn{ s+\Delta s}}.
		\label{infinitesimal evolution}
}
We can define the displacement $D$ vector:
\al{
\fv{\Delta x}
	&:=	\fv x\fn{s+\Delta s}-\fv x\fn{s},
		\label{displacement vector}
}
i.e.,
\al{
\fv{\Delta x}
	=
\bmat{\Delta x^0\\ \Delta\bs x}
	&=	\bmat{
			x^0\fn{s+\Delta s}\\
			\bs x\fn{s+\Delta s}
			}
		-\bmat{
			x^0\fn{s}\\
			\bs x\fn{s}
			}\nn
	&=	\bmat{
			x^0\fn{s+\Delta s}-x^0\fn{s}\\
			\bs x\fn{s+\Delta s}-\bs x\fn{s}
			}
	=	\bmat{
			x^0\fn{s+\Delta s}-x^0\fn{s}\\
			x^1\fn{s+\Delta s}-x^1\fn{s}\\
			\vdots\\
			x^d\fn{s+\Delta s}-x^d\fn{s}
			}.
			\label{explicit displacement vector}
}
There are three possibilities for the particle's move in classical physics:
\begin{itemize}
\item An ordinary \emph{massive} particle always goes time-like: $\lip{\fv{\Delta x}}^2<0$, where the square denotes the Lorentzian norm~\eqref{Lorentzian norm}.
\item A \emph{massless} particle, such as a \emph{photon} of which \emph{light} consists, always goes light-like: $\lip{\fv{\Delta x}}^2=0$.
\item A \emph{tachyon} always goes space-like: $\lip{\fv{\Delta x}}^2>0$.
\end{itemize}
The existence of a tachyon indicates a pathology of the system, and violates \emph{causality} in general.\footnote{\label{tachyon}
In quantum field theory, the existence of a tachyon indicates that one is on a \emph{false vacuum}, which will eventually roll down to a \emph{true vacuum} that does not have a tachyon.}
Therefore, you must be careful in introducing a tachyon in your computer game.
Hereafter, we focus on massive and massless particles.
The time difference in this step is ${\Delta x}^0$.
We assume that none of the massive and massless particles go backward in time.\footnote{
In quantum field theory, a particle going backward in time is identical to its \emph{antiparticle} going forward in time.
}
That is, we assume that ${\Delta x}^0>0$.

For a massive particle, we define its \emph{proper time} $\tau$ such that it increases by an amount
\al{
\Delta\tau
	&:=	\sqrt{-\lip{\fv{\Delta x}}^2}\nn
	&=	\sqrt{\paren{{\Delta x}^0}^2-\paren{\Delta\bs x}^2}
	=	{\Delta x}^0\sqrt{1-\paren{\Delta\bs x\over{\Delta x}^0}^2}
		\label{proper time}
}
for an infinitesimal evolution $\fv{\Delta x}$.
It is important to note that the proper time is, by definition, manifestly \emph{Lorentz invariant}.\footnote{
For massive particles, we may e.g.\ use $\tau$ as the worldline parameter~$s$. Instead, we may also choose an arbitrary monotonically increasing function of $\tau$ as $s$; see footnote~\ref{affine footnote}.
}
Note that the proper time is defined for \emph{each} particle individually.

Now we can define the $D$ velocity for the massive particle:
\al{
\fv u\fn{s}
	&:=	\lim_{\Delta s\to0}
			{\fv x\fn{s+\Delta s}-\fv x\fn{s}\over\Delta\tau}
	=	{\df\fv x\ov \df\tau},
		\label{covariant velocity given}
}
where $\Delta\tau$ is given in Eq.~\eqref{proper time}.
By definition, $\fv u$ is a \emph{Lorentz-covariant} $D$ vector.\footnote{
Something is Lorentz covariant if it transforms as a \emph{representation} of the Lorentz transformations.
In particular, if it transforms as a vector as in Eq.~\eqref{u transformation}, it is a Lorentz-covariant $D$ vector.
}
That is, for a Lorentz transformation $\fv x\to \fv{x'}=\Lambda \fv x$, it is straightforward to show that $\fv u$ is covariant:
\al{
\fv u
	&\to	\fv{u'}
	=	\lim_{\Delta s\to 0}{\Lambda\fv x\fn{s+\Delta s}-\Lambda\fv x\fn{s}\ov\Delta\tau}
	=	\Lambda \fv u,
		\label{u transformation}
}
where we have used the Lorentz invariance of $\Delta\tau$ defined in Eq.~\eqref{proper time}.
Note that, by definition,
\al{
\lip{\fv u}^2
	&=	-1.
		\label{u normalization}
}
We hereafter choose the spatial component $\bs u$ as the $d$ independent parameters, and then the time component of the $D$ velocity is not independent:
\al{
u^0
	&=	\sqrt{1+\bs u^2}
	=	\sqrt{1+\paren{u^1}^2+\cdots+\paren{u^d}^2}.
}
The evolution~\eqref{infinitesimal evolution} for a small proper-time step $\Delta\tau$ now reads
\al{
\fv x\fn{\tau}
	&\longmapsto	\fv x\fn{\tau+\Delta\tau}
			=	\fv x\fn{\tau}+\fv u\fn{\tau}\,\Delta\tau,\nn
\bmat{x^0\fn{\tau}\\ x^1\fn{\tau}\\ \vdots\\ x^d\fn{\tau}}
	&\longmapsto	\bmat{x^0\fn{\tau+\Delta\tau}\\ x^1\fn{\tau+\Delta\tau}\\ \vdots\\ x^d\fn{\tau+\Delta\tau}}
			=	\bmat{x^0\fn{\tau}\\ x^1\fn{\tau}\\ \vdots\\ x^d\fn{\tau}}
				+\bmat{u^0\fn{\tau}\\ u^1\fn{\tau}\\ \vdots\\ u^d\fn{\tau}}
				\Delta\tau.
}

We note that the ordinary Lorentz-noncovariant velocity $\bs v$ in this particular coordinate system is given by
\al{
\bs v
	&=	\lim_{\Delta s\to 0}{\Delta\bs x\over{\Delta x}^0}
	=	{\df\bs x\ov\df t},
		\label{noncovariant velocity}
}
where $\Delta\bs x$ and $\Delta x^0$ are given in Eq.~\eqref{explicit displacement vector} and, in the last step, we have tentatively come back from the natural units for a reader unfamiliar with dimensional analysis; see Eq.~\eqref{0th coordinate}.
Using this Lorentz-noncovariant velocity, one can show that the spatial component of $\fv u$ becomes
\al{
\bs u
	&=	\lim_{\Delta s\to0}{\Delta\bs x\over\sqrt{\paren{{\Delta x}^0}^2-\paren{\Delta\bs x}^2}}
	=	\lim_{\Delta s\to0}{{\Delta\bs x\over{\Delta x}^0}\over\sqrt{1-\paren{\Delta\bs x\over{\Delta x}^0}^2}},
}
i.e.,
\al{
\bs u
	&=	{\bs v\over\sqrt{1-\bs v^2}},	&
\bs v
	&=	{\bs u\over\sqrt{1+\bs u^2}}.
}
There is a one-to-one correspondence between $\bs u$ and $\bs v$.
We see that for a massive particle, the possible values of $\bs v$ are $\ab{\bs v}<1$, or $\ab{\bs v}<c$ when we recover $c$.
On the other hand, $\ab{\bs u}$ can be arbitrarily large.
When the particle's velocity is much smaller than the speed of light $\ab{\bs v}\ll c=1$, we see that the covariant and noncovariant velocities match each other:
\al{
\bs u
	&=	\bs v+\Or{\ab{\bs v}^3},	&
\bs v
	&=	\bs u+\Or{\ab{\bs u}^3},
}
where $\Or{\epsilon^n}$ is the \emph{Landau symbol} representing the terms of order $\epsilon^n$ and higher as $\epsilon\to0$.

From Eq.~\eqref{proper time}, we can write the infinitesimal time difference ${\Delta x}^0$ in this particular coordinate system as
\al{
{\Delta x}^0
	&=	{1\over\sqrt{1-\bs v^2}}\,\Delta\tau.
		\label{time dilation in v}
}
In terms of $\bs u$, the relation~\eqref{time dilation in v} can be written as
\al{
{\Delta x}^0
	&=	\gamma\,\Delta\tau,
		\label{time dilation}
}
where the \emph{Lorentz factor} $\gamma$ is defined by
\al{
\gamma\fn{\bs u}
	&:=	\sqrt{1+\bs u^2}
	=	u^0
	=	{1\over\sqrt{1-\bs v^2}}\geq1.
}
Note that $\gamma\fn{-\bs u}=\gamma\fn{\bs u}$ and that $\gamma\fn{\bs 0}=1$.
Hereafter, $\bs v$ does not play any role in the actual formulation of the relativity since $\bs v$ is a Lorentz-noncovariant quantity, and we simply call $\bs u$ (rather than $\bs v$) the velocity unless otherwise stated.

We also define the $D$ acceleration,
\al{
\fv a\fn{s}
	&:=	\lim_{\Delta s\to0}{\fv u\fn{s+\Delta s}-\fv u\fn{s}\ov\Delta\tau}
	=	{\df\fv u\ov\df\tau},
		\label{acceleration given}
}
where $\Delta\tau$ is given in Eq.~\eqref{proper time}.\footnote{\label{footnote on acceleration}
The spatial component of $D$ acceleration, $\bs a$, differs from the one appearing in the nonrelativistic equation of motion~\eqref{Newtonian eom}: The relativistic and nonrelativistic ones are $\bs a=\df^2\bs x/\df\tau^2=\df u/\df\tau$ and $\bs a_\tx{NR}=\df^2\bs x/\paren{\df x^0}^2=\df\bs v/\df x^0$, respectively. The latter never appears hereafter.
}
Note that, by taking the $\tau$ derivative of $\lip{\fv u}^2=-1$, we obtain
\al{
\fv u\cdot\fv a
	&=	0,
		\label{u dot a}
}
and hence
\al{
a^0
	&=	{\bs u\over\gamma\fn{\bs u}}\cdot\bs a\nn
	&=	{1\over\gamma\fn{\bs u}}\sum_{i=1}^du^ia^i 
	\label{time component of a}
}
in any coordinate system. In particular, $a^0=0$ whenever $\bs u=\bs 0$.

\subsection{Rest frame}\label{rest frame section}
Let us show that we can always take a coordinate system in which a particle appears to be at rest.
Suppose that a particle has a velocity $\bs u$ in a particular coordinate system $\Set{\fv x}$ at a \emph{given single moment} when its proper time is $\tau$.
Then the following Lorentz transformation makes $\fv u\to\paren{1,\bs 0}$:
\al{
L\fn{\bs u}
	&:=	\bmat{
			\gamma\fn{\bs u}&-\bs u^\T\\
			-\bs u  & \I+\paren{\gamma\fn{\bs u}-1}\bh u\bh u^\T
			},
			\label{to rest frame}
}
where $\bh u:=\bs u/\ab{\bs u}$.\footnote{
In $D=1+1$ spacetime dimensions, Eq.~\eqref{to rest frame} becomes
\als{
L=\bmat{\sqrt{1+\paren{u^1}^2}&-u^1\\ -u^1&\sqrt{1+\paren{u^1}^2}},
}
which reduces to the celebrated form appearing in Einstein's original paper~\cite{Einstein:1905ve},
\als{
L=\bmat{{1\ov\sqrt{1-v^2}}&-{v\ov\sqrt{1-v^2}}\\ -{v\ov\sqrt{1-v^2}}&{1\ov\sqrt{1-v^2}}},
}
when written in terms of the noncovariant velocity $v:=u^1/\sqrt{1+\paren{u^1}^2}$.
}
More explicitly,
\al{
\bh u
	&=	{1\over\sqrt{\paren{u^1}^2+\cdots+\paren{u^d}^2}}\bmat{u^0\\ u^1\\ \vdots\\ u^d},&
\bh u\bh u^\T
	&=	{1\over \paren{u^1}^2+\cdots+\paren{u^d}^2}\bmat{
			u^1u^1&\cdots&u^1u^d\\
			\vdots&\ddots&\vdots\\
			u^du^1&\cdots&u^du^d
			}.
}
One can verify that $L\fn{\bs u}$ indeed transforms $\fv u$ to rest, that $L\fn{-\bs u}$ is the inverse matrix of $L\fn{\bs u}$, and that $L\fn{\bs u}$ satisfies the condition for the Lorentz transformation~\eqref{Lorentz condition}, respectively:
\al{
\fv u\to \fv U :=
L\fn{\bs u}\fv u
	&=	\bmat{1\\ \bs 0}, &
L\fn{-\bs u}
	&=	\sqbr{L\fn{\bs u}}^{-1}, &
\sqbr{L\fn{\bs u}}^\T\eta\,L\fn{\bs u}
	&=	\eta.
}

In the new coordinate system $\Set{\fv X|\fv X=L\fn{\bs u}\fv x}$, namely the \emph{rest frame} for the particle at the proper time $\tau$, we have $\bs U=\bs 0$ at that moment. 
Hereafter, we write quantities in someone's rest frame in upper case in general, unless otherwise stated.
During the subsequent infinitesimal time evolution, we get $\Delta\tau=\Delta X^0$ because $\gamma\fn{\bs 0}=1$. That is, the proper-time flow of the particle is nothing but the time flow in each rest frame in each time step. Therefore, the proper time is the time felt by the particle itself.
Now we can interpret from Eq.~\eqref{time dilation} that the time difference ${\Delta x}^0$ in an arbitrary frame (other than the rest frame) is always larger than the proper time $\Delta\tau$ felt by the moving particle itself.\footnote{
The time difference ${\Delta x}^0$ is not an actual time difference seen by any observer.
We will see that the actual time difference for an observer is the one deduced from the time foliation by the observer's past light cones rather than by an equal-time slice. See Sec.~\ref{formulation}. 
}
This phenomenon is called \emph{time dilation}.

Einstein's \emph{equivalence principle} asserts that the physical law for a moving particle is the same as the one that is Lorentz-transformed from the law of the particle at rest. 
That is, one can write the relativistic equation of motion just by replacing $\df t$ by $\df\tau$:\footnote{
As both sides are \emph{covariant}, the equation of motion becomes \emph{invariant} under the Lorentz transformations: $m\fv a=\fv f$ is transformed to $m\Lambda\fv a=\Lambda\fv f$, which comes back to $m\fv a=\fv f$ by multiplying $\Lambda^{-1}$ on the both sides.
}
\al{
m\fv a
	&=	\fv f,
		\label{equation of motion}
}
where $\fv f$ is nothing but the $D$ vector transformed from the rest-frame force felt by the particle:
\al{
\fv f=L\fn{-\bs u}\bmat{0\\ \bs F}.
	\label{D fource}
}
Note that the 0th (time) component of the $D$ force in the rest frame must be zero in order for the equation of motion~\eqref{equation of motion} to hold; see the discussion after Eq.~\eqref{time component of a}.
That is, the 0th component of the equation of motion does not have independent information.

Conversely, for a given acceleration $\bs A$ in the rest frame, we obtain the $D$ acceleration in any frame in which the particle has the velocity $\bs u$:
\al{
\fv{a}
	&=	L\fn{-\bs u}\bmat{0\\ \bs A}
	=	\bmat{\bs u\cdot\bs A\\ \bs A+\paren{\gamma\fn{\bs u}-1}\paren{\bs A\cdot\bh u}\bh u}.
}
That is,
\al{
\bs a=\bs A+\paren{\gamma\fn{\bs u}-1}\paren{\bs A\cdot\bh u}\bh u.
}
In particular, if there is no acceleration in the rest frame $\bs A=\bs 0$, the acceleration vanishes in any frame $\bs a=\bs 0$.
We may also show that, when $\bs a=\bs 0$ in a frame, then the $D$ acceleration vanishes in any frame.\footnote{\label{acceleration footnote}
Suppose that $\bs a=\bs 0$ in a frame: $\fv a=\paren{a^0,\bs 0}$. We write the velocity in this frame $\bs u$. Then the rest-frame acceleration reads
\als{
\bmat{0\\ \bs A}
	&=
L\fn{\bs u}\bmat{a^0\\ \bs 0}
	=	a^0\bmat{\gamma\fn{\bs u}\\ -\bs u}.
}
Since $\gamma\fn{\bs u}\geq1$, $a^0=0$ results from the time component of this equation. This means $\fv a=0$. Any Lorentz transformation of the zero vector is zero: $\Lambda\fv 0=\fv 0$. Therefore the $D$ acceleration is zero in any frame if $\bs a=\bs 0$ in a frame.
}

We note that it is perfectly legitimate to work within the framework of special relativity in order to handle an acceleration unless it is due to gravitational interactions. When and only when the gravitational interactions are strong enough, we need general relativity for a full treatment.

\section{Drawing the world}\label{implementation}
\begin{figure}[tp]
\begin{center}
\includegraphics[width=0.6\textwidth]{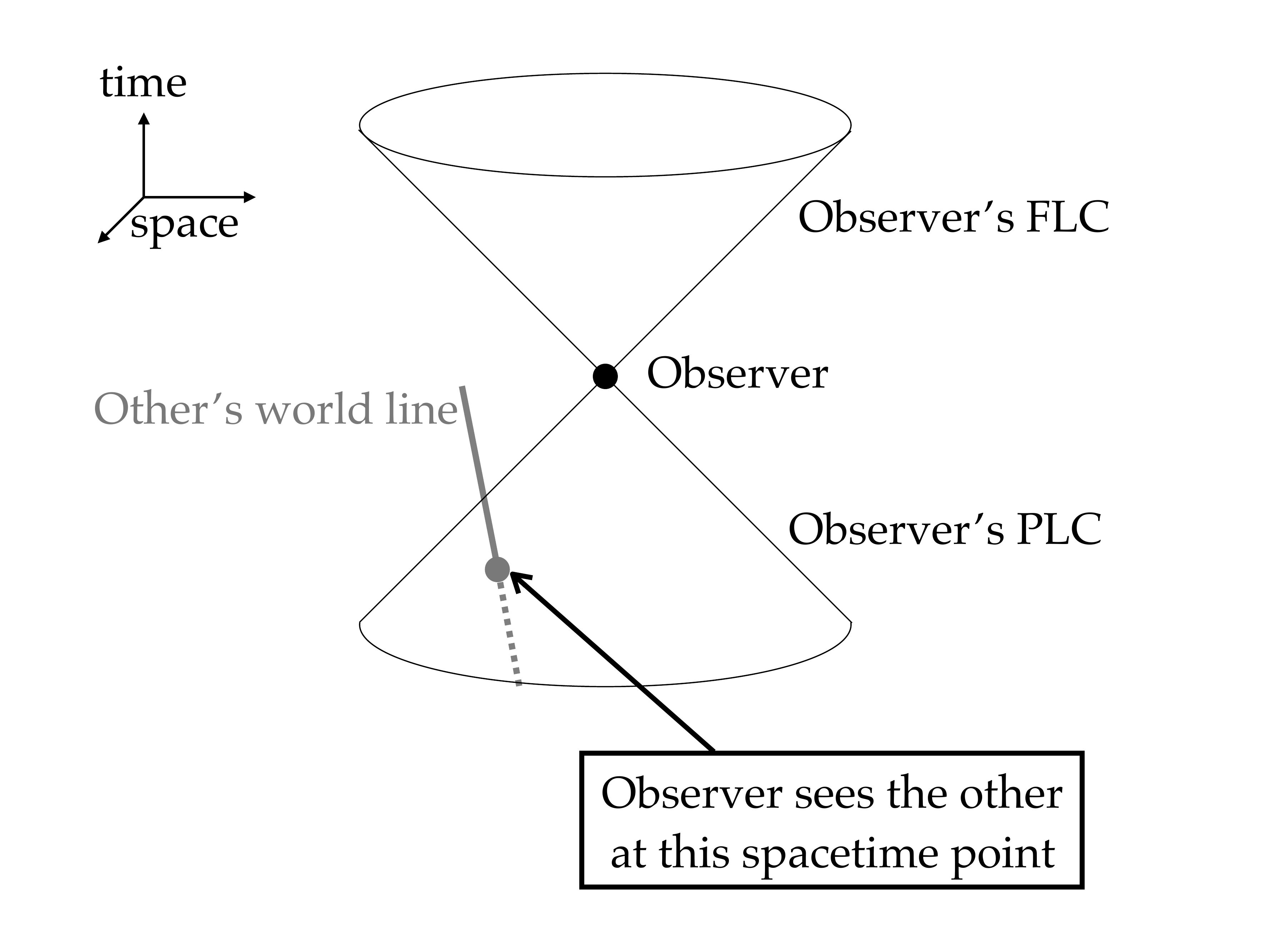}
\caption{Schematic figure showing the observer's future and past light cones in $2+1$ spacetime dimensions. The other's worldline is shown to indicate the spacetime point that is seen by the observer.}\label{schematic figure first}
\end{center}
\end{figure}

We show how to draw the relativistic world for a given set of worldlines of all the objects. 
A schematic figure for this section is presented in Fig.~\ref{schematic figure first}.

First we arbitrarily choose a \emph{reference frame} $\Set{\fv x}$ and store all the information as written in this coordinate system. 
In practice, it is convenient to choose a reference frame such that background objects are at rest in the frame.\footnote{
In our real Universe, we can define the absolute rest frame such that the dipole component of the cosmic microwave background vanishes; see Ref.~\cite{Adam:2015rua} for latest observational results on the cosmic microwave background.
}
Hereafter, we write the reference-frame quantities in lower case, unless otherwise stated. 

\subsection{Drawing the world on the observer's past light cone}\label{formulation}
For an observer at a spacetime point $\fv\xO$, the hypersurface consisting of possible light rays that can come to $\fv\xO$ is the past light cone (PLC):
\al{
\PLC\fn{\fv\xO}
	&=	\Set{\fv x|\lip{\fv x-\fv\xO}^2=0,\quad x^0<\xO^0}.
}
For later use, we also define the future light cone (FLC):
\al{
\FLC\fn{\fv\xO}
	&=	\Set{\fv x|\lip{\fv x-\fv\xO}^2=0,\quad x^0>\xO^0}.
}

We say that a spacetime point~$\fv x$ is on the \emph{past side} of $\PLC\fn{\fv \xO}$ when $\lip{\fv x-\fv \xO}^2<0$ (time-like separation) and $\xO^0>x^0$ ($\fv x$ is on the past of $\fv\xO$).
We also say that $\fv x$ is on the \emph{future side} of $\PLC\fn{\fv \xO}$ when $\fv x$ is neither on $\PLC\fn{\fv \xO}$ nor on its past side.
The observer can see the world sliced by its $\PLC\fn{\fv \xO}$, i.e., the observer can know things on $\PLC\fn{\fv \xO}$ and its past side. 
It is important that the PLC is Lorentz invariant in the sense that, if a particle is on the PLC, it remains so in any Lorentz-transformed frame.\footnote{
More precisely, this is the case for the \emph{proper orthochronous Lorentz transformation} that satisfies $\det\Lambda=1$ and $\Lambda^0{}_0\geq1$, where $\det\Lambda=\pm1$ and $\ab{\Lambda^0{}_0}\geq1$ follow from Eq.~\eqref{Lorentz condition} generally. Any Lorentz transformation is either proper orthochronous or obtained by multiplying a \emph{time reversal} $\diag\fn{-1,1,1,1}$ and/or a \emph{space inversion} $\diag\fn{1,-1,-1,-1}$ onto the proper orthochronous one.
}
This is also the case for the notion of the past and future sides of the PLC, and similarly for the FLC and both its sides.

Let $W_n$ be the worldline of the $n$th particle:
\al{
W_n	&=	\Set{\fv{x_n}\fn{\tau}|\tau_n^\tx{ini}\leq\tau\leq\tau_n^\tx{fin}},
}
where $\tau_n^\tx{ini}$ and $\tau_n^\tx{fin}$ are the initial and final proper times of the worldline.
The observer at $\fv\xO$ sees this particle at the intersecting point between $\PLC\fn{\fv \xO}$ and $W_n$, which we write as
\al{
\fv{x_n}^{\PLC\paren{\fv\xO}}
}
hereafter.
How to determine the intersection in a concrete implementation will be explained in Sec.~\ref{sec:intersection}.

An observer having a velocity $\bs \uO$ in the reference frame $\Set{\fv x}$ sees the world in the \emph{observer's rest frame}:\footnote{\label{central frame footnote}
If one wants to place the observer at the origin, one may take, so to say, a \emph{central frame} $\Set{\fv X|\fv X=L\fn{\bs\uO}\paren{\fv x-\fv\xO}}$, which differs from the rest frame~\eqref{rest frame} just by a \emph{translation}, a constant shift of the coordinate origin by the $D$ vector $L\fn{\bs\uO}\fv\xO$; see footnote~\ref{Poincare}.
}
\al{
\Set{\fv X|\fv X=L\fn{\bs\uO}\fv x}.
	\label{rest frame}
}
Each of all the other particles, say the $n$th one, is seen by the player as located at
\al{
\fv{X_n}^{\PLC\paren{\fv{X_\tx{O}}}}=L\fn{\bs u_{\tx O}}\fv{x_n}^{\PLC\paren{\fv \xO}},
	\label{new location}
}
where $\fv{X_\tx{O}}=L\fn{\bs \uO}\fv{\xO}$.\footnote{
$\PLC\fn{\fv{\xO}}$ and $\PLC\fn{\fv{X_\tx{O}}}$ are the same. We use the latter in the left-hand side of Eq.~\eqref{new location} so that all quantities are given in the rest frame there.
}
Similarly, the velocity $\fv{u_n}^{\PLC\paren{\fv \xO}}$ of the $n$th particle at $\fv{x_n}^{\PLC\paren{\fv \xO}}$ is seen by the player as
\al{
\fv{U_n}^{\PLC\paren{\fv{X_\tx{O}}}}=L\fn{\bs u_{\tx O}}\fv{u_n}^{\PLC\paren{\fv \xO}}.
}
These are all needed to draw the world fully relativistically.

\subsection{Discrete worldline}\label{sec:worldline}
The program stores the worldline of each particle in the reference frame~$\Set{\fv x}$.
Due to the iterations, each worldline becomes a discrete set of its past spacetime points, just as in the Newtonian case~\eqref{Newtonian discrete worldline}:
\al{
W	&=	\Set{\fv x\fn{s_1},\,\fv x\fn{s_2},\,\dots,\,\fv x\fn{s_{N_W}}}
	=:	\Set{\fv{x_1},\dots,\fv{x_{N_W}}},
	\label{discrete worldline}
}
where 
$s_\ttt$ ($\ttt=1,\dots,N_W$) is the parameter of the worldline, which can be identified as the proper time of the particle---we have abbreviated them as $\fv{x_\ttt}:=\fv x\fn{s_\ttt}$---and 
we always order from past to future: $s_1<s_2<\cdots<s_{N_W}$.
The velocity $\fv{u_\ttt}$ between $\fv{x_{\ttt-1}}$ and $\fv{x_\ttt}$ is given as
\al{
\fv{u_\ttt}
	&=	{\fv{\Delta x_\ttt}\over\sqrt{-\sqbr{\fv{\Delta x_\ttt}}^2}},
		\label{eq:u_j}
}
where $\fv{\Delta x_\ttt}:=\fv{x_\ttt}-\fv{x_{\ttt-1}}$.

\subsection{Intersection between worldline and PLC}\label{sec:intersection}

As discussed above, it is important to compute the intersection between a worldline and a light cone.
Let us spell out the method to obtain the intersection between the worldline~\eqref{discrete worldline} and $\PLC\fn{\fv{\xO}}$:
\begin{itemize}
\item From the past to the future $\fv{x_\ttt}$ with $\ttt=1,2,\dots$, we check if $\fv{x_\ttt}$ is on the past side of $\PLC\fn{\fv{\xO}}$; namely, check if
\al{
\lip{\fv{\xO}-\fv{x_\ttt}}^2<0\qquad \tx{and} \qquad \xO^0>x_\ttt^0.
}
The first point $\fv{x_{\ttt'}}$ that violates this condition is the point that is closest to $\PLC\fn{\fv{\xO}}$ in $W$ on the future side of $\PLC\fn{\fv{\xO}}$.
\item Between the adjacent points $\fv{x_{\tt t'-1}}$ and $\fv{x_{\ttt'}}$, the worldline is obtained by linear interpolation:
\al{
\fv x\fn{\sigma}
	&=	\paren{1-\sigma}\,\fv{x_{\tt t'-1}}+\sigma\,\fv{x_{\ttt'}},
}
where $0\leq \sigma\leq 1$.
\item The intersecting point $\fv{x}\fn{\sigma}$ should satisfy
\al{
\lip{\fv x\fn{\sigma}-\fv{\xO}}^2=0,
}
i.e.,\footnote{\label{FLC intersection}
The other point $\sigma=\paren{\beta+\sqrt{\beta^2-\alpha\gamma}}/\alpha$ gives the intersection with the future light cone.
}
\al{
\sigma	&=	{\beta-\sqrt{\beta^2-\alpha\gamma}\over \alpha},
}
where\footnote{
be careful with the abuse of notation: $\gamma$ here has nothing to do with the Lorentz factor.
}
\al{
\alpha	&:=	-\lip{\fv{\xO}-\fv{x_{\ttt'-1}}}^2>0,\\
\beta	&:=	-\lip{\fv{\xO}-\fv{x_{\ttt'-1}},\,\fv{x_{\ttt'}}-\fv{x_{\ttt'-1}}}>0,\\
\gamma	&:=	-\lip{\fv{x_{\ttt'}}-\fv{x_{\ttt'-1}}}^2>0.
}
This value of $\sigma$ gives the intersecting point $\fv x^{\PLC\paren{\fv{\xO}}}$.
\end{itemize}

\subsection{Short summary}
At each moment, we draw all the objects as if each of them is placed at the spatial position $\bs x^{\PLC\paren{\fv{\xO}}}$.
Once all the spatial positions are given, concrete implementation of the drawing of the world is the same as in ordinary nonrelativistic 3D games.
An approximate treatment of a rigid body will be presented in Sec.~\ref{rigid body section}.

\section{Time evolution}\label{time evolution}
\begin{figure}[tp]
\begin{center}
\includegraphics[width=0.7\textwidth]{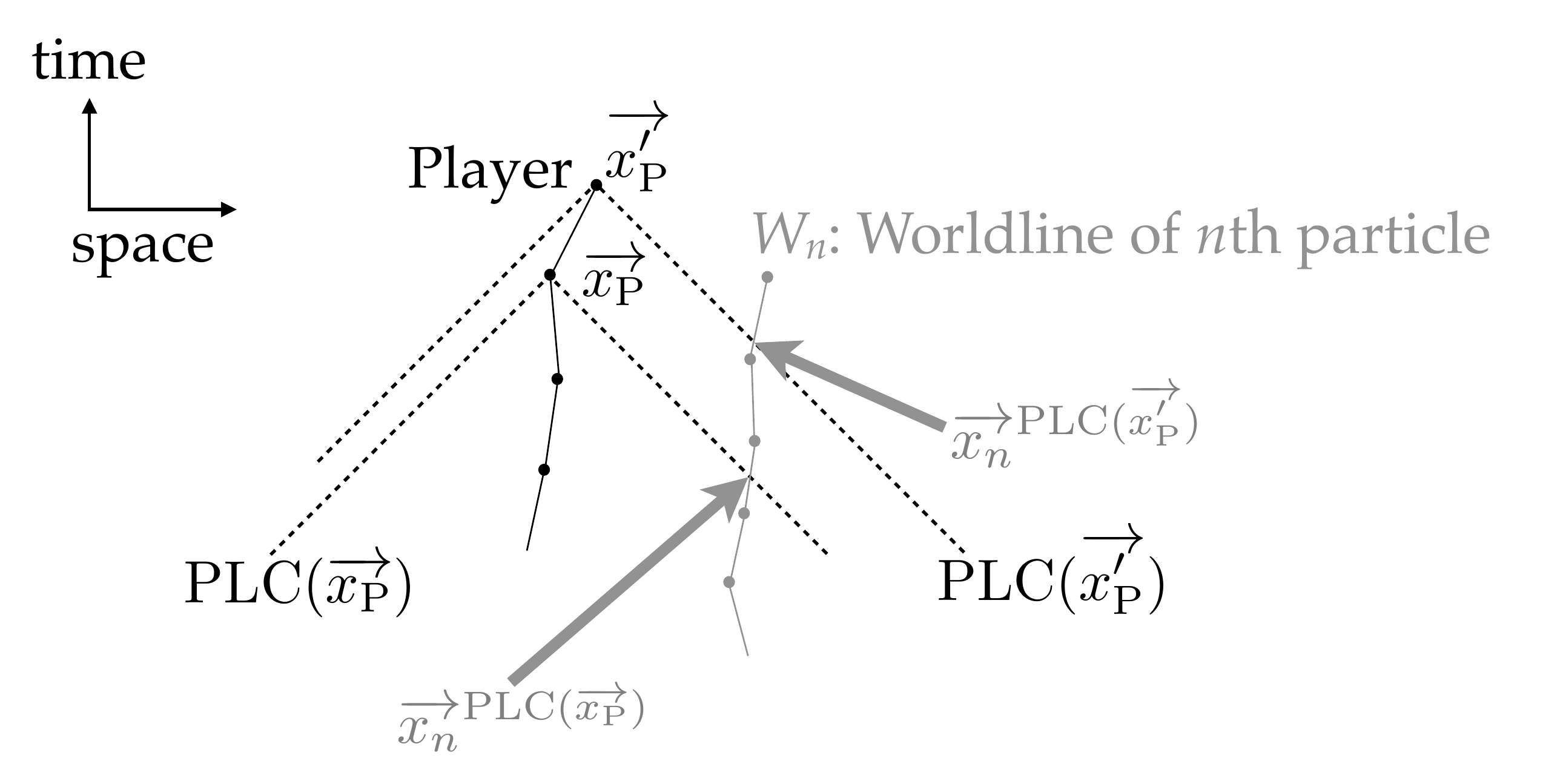}
\caption{Schematic diagram in $1+1$ spacetime dimensions for the time evolution of the player and the $n$th particle under the PLC foliation.}
\label{time evolution figure}
\end{center}
\end{figure}

We show how to take into account the relativistic time evolution of the system, with an application to a first-person shooter (FPS) in mind.
As reviewed above, a game in Newtonian mechanics draws the world at each \emph{equal-time slice}. One problem is that such a time slice is not Lorentz invariant and is not compatible with relativity. Instead we employ the foliation of spacetime by the PLCs of the player, which is Lorentz invariant, as reviewed above. A schematic diagram is shown in Fig.~\ref{time evolution figure}.

\subsection{Player's time evolution}\label{player's time evolution}

What should the time evolution be?
At each time step, the future-most surface of the world that is perceivable by the player is $\PLC\fn{\fv \xP}$, on which  the particle's latest position $\fv x$ is located. In the next iteration, the game program calls the real time passed, $\Delta t$, which is identified as the player's proper time passed, $\Delta\tauP$. In the iteration, the player's proper time, position, and velocity move to, respectively,
\al{
\tauP
	&\longmapsto	\tauP+\Delta\tauP,\nn
\fv\xP
	&\longmapsto	\fv\xP+\fv\uP\,\Delta\tauP,\nn
\fv\uP
	&\longmapsto	\fv\uP+\fv\aP\,\Delta\tauP,\label{updating position etc}
}
with
\al{
\fv\aP
	&:=	L\fn{-\bs u_\P}\bmat{0\\ \bs \AP},
}
where $\bs \AP$ is the acceleration of the player given at its rest frame, determined by
\al{
\bs \AP={\bs F_\P\over m_\P},
}
in which $\bs F_\P$ is the ordinary Newtonian force felt by the player in its rest frame, coming both from the player's own input and from influences from its environment and others.
Concrete examples of how to give $\bs\AP$ will be shown in Sec.~\ref{sec:move}.

We update the values of the player's proper time, position, and velocity as in Eq.~\eqref{updating position etc}.
Now the player's new position $\fv{\xP'}$ is determined, and we know the new $\PLC\fn{\fv{\xP'}}$ which is used to draw the world in the next iteration. 

We add the new point
\al{
\fv{\xP'}:=\fv\xP+\fv\uP\Delta\tauP
}
to the last of the player's worldline set, as described in Sec.~\ref{sec:worldline}.
The player may have also done some action at $\fv\xP$ before its move, e.g., have shot a beam from $\fv\xP$, which will be treated later in Sec.~\ref{shooting section}.

\subsection{Others' time evolution}\label{others' time evolution}

The $n$th particle at $\fv{x_n}^{\PLC\paren{\fv \xP}}$ determines its own move, and extends its worldline $W_n$, according to its own acceleration and influences from others, until the extended $W_n$ hits the new $\PLC\fn{\fv{\xP'}}$.
The new intersection point is drawn in the next iteration as above.

Now we show how to determine the moves of NPCs. Suppose that an NPC, say the $n$th one, is at $\fv{x_n}^{\PLC\paren{\fv \xP}}$, and has a velocity $\fv u$ and  a proper time~$\tau$, all in the reference frame. The NPC can see the world sliced by its own $\PLC\fn{\fv{x_n}}$, or, more precisely, can know things on $\PLC\fn{\fv{x_n}}$ and its past side. According to the available information, we determine the NPC's acceleration $\fv{a_n}$ in the reference frame; see Sec.~\ref{sec:move} for details.
The NPC's move is
\al{
\tau_n
	&\longmapsto	\tau_n+\Delta\tau_n,\\
\fv{x_n}	
	&\longmapsto	\fv{x_n}+\fv{u_n}\,\Delta\tau_n,\\
\fv{u_n}	
	&\longmapsto	\fv{u_n}+\fv{a_n}\,\Delta\tau_n,
	\label{velocity evolution}
}
where $\Delta\tau_n$ can be chosen to be $\Delta\tauP$, or else whatever (fixed or variable) value according to the NPC's reflexes. In general, the resultant $\fv{x_n}+\fv{u_n}\,\Delta\tau_n$ is still on the past side of $\PLC\fn{\fv{\xP'}}$. We iterate until it goes beyond $\PLC\fn{\fv{\xP'}}$, namely, until the following condition is violated:
\al{
\lip{\fv{\xP'}-\fv{x_n}}^2<0\qquad\tx{and}\qquad \xP^{\pr0}>x_n^0.
}

In each step, the NPC may also do some action, as will be described in Secs.~\ref{shooting section}.
Note that the NPC at $\fv{x_n}$ sees the world cut by its own $\PLC\fn{\fv{x_n}}$, and cannot see further moves of others beyond it: Any other NPC, say the $m$th one, is seen by the one at $\fv{x_n}$ as located at the intersecting point between $W_m:=\Set{\fv{x_m}\fn{\tau_m}|\text{all $\tau_m$}}$ and $\PLC\fn{\fv{x_n}}$, which we write as $\fv{x_m}^{\PLC\paren{\fv{x_n}}}$. To repeat, the one at $\fv{x_n}$ can know things only on the past side of $\PLC\fn{\fv{x_n}}$.

\subsection{Aiming}\label{sec:aiming}
\begin{figure}[tp]
\begin{center}
\includegraphics[width=0.7\textwidth]{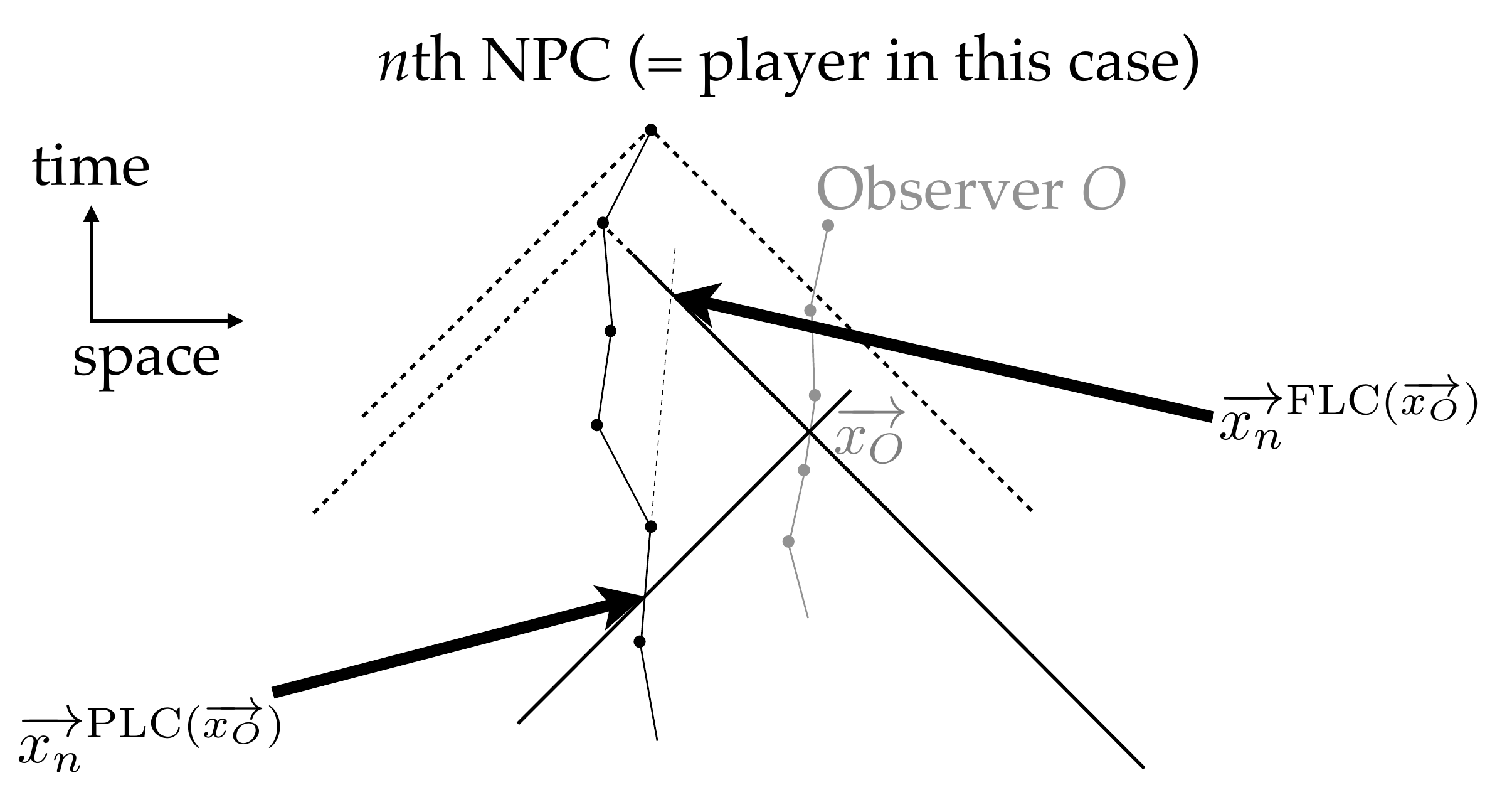}
\caption{Schematic diagram in $1+1$ spacetime dimensions for the aiming.}
\label{FLC figure}
\end{center}
\end{figure}

We show how an observer $O$ can estimate the future position of other objects.
A schematic diagram is shown in Fig.~\ref{FLC figure}.
The observer at $\fv{\xO}$ sees the world sliced by its own $\PLC\fn{\fv{\xO}}$, on which the perceivable future-most position of another object, say the $n$th NPC,\footnote{
Here this ``$n$th NPC'' may indicate the player too.
}
is located as $\fv{x_n}^{\PLC\paren{\fv{\xO}}}$.
From the given information, $O$ wants to predict the $n$th NPC's future move.

Let $\fv{u_n}$ be its velocity at $\fv{x_n}^{\PLC\paren{\fv{\xO}}}$.
We want to know the intersection between the FLC of $O$,
\al{
\FLC\fn{\fv{\xO}}
	&:=	\Set{\fv{x}|\lip{\fv x-\fv{\xO}}^2=0,\quad  x^0>\xO^0},
}
and the would-be worldline of the $n$th NPC extended in the direction of $\fv{u_n}$:
\al{
\Set{\fv{x_n}+\tau\,\fv{u_n}|\tau>0}.
	\label{n's would-be worldline}
}
The value of $\tau$ at the intersection is obtained from
\al{
0	&=	\lip{\fv{x_n}+\tau\,\fv{u_n}-\fv{\xO}}^2\nn
	&=	\lip{\fv{x_n}-\fv{\xO}}^2+2\tau\lip{\fv{x_n}-\fv{\xO},\,\fv{u_n}}+\tau^2\lip{\fv{u_n}}^2\nn
	&=	2\tau\lip{\fv{x_n}-\fv{\xO},\,\fv{u_n}}-\tau^2,
}
where we have used the PLC condition $\lip{\fv{x_n}-\fv{\xO}}^2=0$ and Eq.~\eqref{u normalization} in the last step. That is, the projected location of the $n$th NPC onto $O$'s FLC is given by
\al{
\tau	&=	2\lip{\fv{x_n}-\fv{\xO},\,\fv{u_n}},
}
namely, at
\al{
\fv{x_n}^{\FLC\paren{\fv{\xO}}}
	&:=	\fv{x_n}+2\lip{\fv{x_n}-\fv{\xO},\,\fv{u_n}}\,\fv{u_n}.
		\label{n's future projection}
}
This point is nothing but the intersection with the FLC mentioned in footnote~\ref{FLC intersection}.
The observer will see this coordinate in its rest frame at
\al{
\fv{X_n}^{\FLC\paren{\fv{\xO}}}
	&:=	L\fn{\fv{\uO}}\,\fv{x_n}^{\FLC\paren{\fv{\xO}}}.
		\label{future projection in rest frame}
}

\subsection{Shooting}\label{shooting section}
We continue the discussion from the previous subsection.
Now $O$ wants to shoot a beam or bullet, $B$, at the $n$th NPC from the location $\fv{\xO}$, when its velocity is $\fv{\uO}$.

In $O$'s rest frame $\Set{X}$, we define the ``future beam/bullet cone'' (FBC):
\al{
\FBC\fn{\fv{X_\tx{O}}}
	&:=	\Set{\fv{X_\tx{O}}+s\fv N|\fv N=\bmat{V^{-1}\\ \bh D},\quad\ab{\bh D}=1,\quad s\geq0},
}
where $\fv{X_\tx{O}}:=L\fn{\bs \uO}\fv{\xO}$ is the location of $O$ in its rest frame,
$V$ is the speed of $B$ satisfying $0<V\leq1$, and
$\bh D$ is a possible direction in which $B$ can be shot.

In $O$'s rest frame, the would-be worldline of $n$th NPC~\eqref{n's would-be worldline} becomes
\al{
\Set{\fv{X_n}+\tau\fv{U_n}|\tau>0},
	\label{n's would-be worldline in rest frame}
}
where $\fv{X_n}:=L\fn{\bs \uO}\fv{x_n}$ and $\fv{U_n}:=L\fn{\bs \uO}\fv{u_n}$.
Its intersection with $\FBC\fn{\fv{X_\tx{O}}}$, if it exists, is given by $\fv{X_\tx{O}}+s\fv N=\fv{X_n}+\tau\fv{U_n}$, i.e.,
\al{
X_\tx{O}^0+{s\over V}
	&=	X_n^0+\tau U_n^0,
		\label{time component of FBC eq}\\
\bs X_\tx{O}+s\bh D
	&=	\bs X_n+\tau\bs U_n.
}
From Eq.~\eqref{time component of FBC eq}, we get
\al{
s	&=	V\paren{X_n^0-X_\tx{O}^0+\tau U_n^0},
}
and hence
\al{
\lip{\fv{X_n}-\fv{X_\tx{O}}+\tau\fv{U_n}}_V^2
	&=	0,
		\label{tau eq}
}
where we have defined the following (linear and bilinear) operations:
\al{
\lip{\fv A}_V^2
	&:=	-V^2\paren{A^0}^2+\bs A^2,	
		\label{bullet norm}\\
\lip{\fv A,\fv B}_V
	&:=	-V^2A^0B^0+\bs A\cdot\bs B.
}
Note that this operation is not Lorentz invariant.
Solving Eq.~\eqref{tau eq} as in Sec.~\ref{sec:intersection}, we get\footnote{
Here $\sqbr{\fv A,\fv B}_V^2$ is the square of $\sqbr{\fv A,\fv B}_V$. Do not confuse it with the norm~\eqref{bullet norm}: the distinction is made by the comma inside.
}
\al{
\tau
	&=	{\lip{\fv{X_n}-\fv{X_\tx{O}},\,\fv{U_n}}_V+\sqrt{\lip{\fv{X_n}-\fv{X_\tx{O}},\,\fv{U_n}}_V^2-\lip{\fv{U_n}}_V^2\lip{\fv{X_n}-\fv{X_\tx{O}}}_V^2}\over -\lip{\fv{U_n}}_V^2}\nn
	&=:	\tau_\tx{F}.
		\label{tau result}
}
It is possible that the would-be worldline~\eqref{n's would-be worldline in rest frame} does not intersect with $\FLC\fn{X_n}$, in which case the expression inside the square root in Eq.~\eqref{tau result} becomes negative.
In such a case, $O$ may still shoot in the direction given in Sec.~\ref{sec:aiming}, which always exists.

$O$ can shoot a beam/bullet whose worldline is given in $O$'s rest frame by
\al{
\Set{\fv{X_\tx{O}}+s\fv{N_\tx{F}}|s\geq0}
	\label{B worldline in rest frame}
}
where
\al{
\fv{N_\tx{F}}
	&:=	\bmat{V^{-1}\\ \bh D_\tx{F}},
}
in which
\al{
\bh D_\tx{F}
	&:=	{1\over V}{\bs X_n-\bs X_\tx{O}+\tau_\tx{F}\bs U_n\over X_n^0-X_\tx{O}^0+\tau_F U_n^0}.
}
In the reference frame, the beam/bullet worldline~\eqref{B worldline in rest frame} becomes
\al{
\Set{\fv{\xO}+s\,\fv{n_\tx{F}}|s\geq0},
}
where
\al{
\fv{\xO}
	&:=	L\fn{-\bs \uO}\fv{X_\tx{O}},\\
\fv{n_\tx{F}}
	&:=	L\fn{-\bs \uO}\fv{N_\tx{F}}.
}

\subsection{Acceleration}\label{sec:move}
Now we exhibit how $O$'s acceleration, used in Sec.~\ref{others' time evolution}, is determined.
There can be several sources for $O$'s acceleration:
\begin{itemize}
\item To match ordinary human common sense, one may suppose that the spacetime is filled with (fictitious) air or fluid that is static in the reference frame, and may assume a friction force that is a function of the velocity of each particle in the reference frame:
\al{
\fv{a_\tx{f}}
	&=	\bmat{a_\tx{f}^0\\ -f\fn{\bs \uO}\bh \uO},
}
where $f\fn{\bs u}>0$ can be any function of $\bs u$, which is typically $\propto\ab{\bs u}$ and $\bs u^2$ for the \emph{laminar} and \emph{turbulent drags}, respectively. Note that the time component can always be obtained from $\fv{a_\tx{f}}\cdot\fv{\uO}=0$:
\al{
a_\tx{f}^0
	&=	{\bs a_\tx{f}\cdot\bs \uO\over \uO^0}
	=	-f\fn{\bs \uO}{\ab{\bs \uO}\over \uO^0}.
}
\item $O$ can accelerate in a direction that $O$ wants. That is, $O$ may self-accelerate by a magnitude $A_\tx{s}$ in a direction $\bh D$ in its rest frame. 
Then $O$'s self-acceleration in its rest frame is given by
\al{
\fv{A_\tx{s}}
	&=	\bmat{0\\ A_\tx{s}\,\bh D},
		\label{self acceleration}
}
and in the reference frame by
\al{
\fv{a_\tx{s}}
	&=	L\fn{-\bs \uO}\fv{A_\tx{s}}.
		\label{constant acceleration}
}
See also Appendix~\ref{rocket propulsion} for a more elaborate rocket propulsion.

For example, $O$ may want to accelerate in the projected direction $\fv d$ of the $n$th particle on $O$'s FLC given by Eq.~\eqref{n's future projection}:
\al{
\fv d
	&:=	\fv{x_n}^{\FLC\paren{\xO}}-\fv{\xO}.
}
The direction of acceleration, $\bh D=\bs D/\ab{\bs D}$, in $O$'s rest frame can be obtained from the spatial component $\bs D$ of the vector
\al{
\fv D
	&:=	L\fn{\bs \uO}\fv d.
}

\item We show how to implement a collision, namely, a repulsive force exerted by the $n$th particle onto $O$, without contradicting causality.
Suppose that the $n$th particle is on $\PLC\fn{\fv{\xO}}$, and we write its and $O$'s locations as $\fv{X_n}$ and $\fv{X_\tx{O}}$, respectively, in the $n$th particle's rest frame. We write the corresponding velocities as $\fv{U_n}=\bmat{1\\ \bs 0}$ and $\fv{U_\tx{O}}=L\fn{\bs u_n}\fv{\uO}$, respectively.
We write the displacement vector from $n$ to $O$ in $n$'s rest frame as:
\al{
\fv{X_{n\to O}}
	&:=	\fv{X_\tx{O}}-\fv{X_n}.
}
We assume that the $n$th particle exerts a strong repulsive force $\fv{F_{n\to O}}$ when the distance in its rest frame is smaller than its typical radius~$R_n$:
\al{
\fv{F_{n\to O}}
	&=	\bmat{F_{n\to O}^0\\ \bs F_{n\to O}^0},
}
where
\al{
\bs F_{n\to O}
	&=	F\,\theta\fn{R_n-\ab{\bs X_{n\to O}}}\,\bh X_{n\to O},
}
in which $F>0$ is a constant and
\al{
\theta\fn{x}
	&=	\begin{cases}
		1	&	(x>0),\\
		0	&	(x<0),
		\end{cases}
}
is the Heaviside step function. If this repulsive force is the only force acting on $O$, it is proportional to $O$'s acceleration $\fv{F_{n\to O}}\propto \fv{A_\tx{O}}$, and hence the condition $\fv{A_\tx{O}}\cdot\fv{U_\tx{O}}=0$ determines the time component of $\fv{F_{n\to O}}$:
\al{
F_{n\to O}^0
	&=	{\bs U_\tx{O}\cdot\bs F_{n\to O}\over U_\tx{O}^0}.
}
Then the acceleration from this repulsive force is, in the reference frame,
\al{
\fv{a_{n\to O}}
	&=	{L\fn{-\bs u_n}\fv{F_{n\to O}}\over m_\tx{O}}.
}
We sum up $\fv{a_{n\to O}}$ coming from all the particles on $\PLC\fn{\fv{\xO}}$.
\end{itemize}
In the end, $O$'s total acceleration is given by the vector sum
\al{
\fv{a_\tx{O}}
	&=	\fv{a_\tx{f}}+\fv{a_\tx{s}}+\sum_{n\tx{ on }\PLC\paren{\xO}}\fv{a_{n\to O}},
}
which is then put into Eq.~\eqref{velocity evolution}.

\section{Miscellaneous ideas}\label{miscellaneous}
We show various ideas to make more realistically relativistic scenes.

\subsection{Faraway scenes on the skydome}\label{skydome}
One puts a sufficiently large sphere that surrounds the player in the player's rest frame, and draws a background texture of the faraway scene on it: At an angle $\paren{\Theta,\Phi}$ on the sphere, we draw the corresponding texture at $\paren{S,T}$ in the texture coordinates. Let us see how $\paren{S,T}$ is determined for a given direction $\paren{\Theta,\Phi}$.

Let $\Set{\fv x}$ be the reference frame in which faraway background objects are at rest.
Suppose that the player and a background object $O$ are at $\fv\xP$ and $\fv\xi$, respectively, in this frame.
In this subsection, we parametrize $O$'s position by
\al{
\fv\xi	&=	\paren{t,x,y,z}=\bmat{t\\ x\\ y\\ z}.
}
Using ordinary polar coordinates, the intersection between $O$'s worldline and the player's PLC can be parametrized as
\al{
\fv\xi	&=	\bmat{-r\\ r\sin\theta\cos\phi\\ r\sin\theta\sin\phi\\ r\cos\theta},
}
where we have assumed that $O$ is so far away that player's position $\fv \xP$ can be regarded as being at the spacetime origin: $\fv\xi-\fv\xP\approx \fv\xi$.
Conversely, we may write the zenith and azimuthal angles as
\al{
\theta
	&=	\arctan{\sqrt{x^2+y^2}\over z},	&
\phi
	&=	\arctan{y\over x}.
		\label{theta phi}
}
We may use texture mapping of the background image using the \emph{equirectangular projection}.
For example, in OpenGL, the background scene at the angle $\paren{\theta,\phi}$ in the reference frame corresponds to the texture coordinates $(S,T)$:
\al{
S	&=	{\phi\over2\pi},	&
T	&=	1-{\theta\over\pi},
	\label{ST vs theta phi}
}
where $0\leq S\leq 1$ and $0\leq T\leq 1$.

Let us consider light that comes into the player's eyes from the $\paren{\Theta,\Phi}$ direction in the player's rest frame.
Let $\fv\Xi$ be the position of the faraway object seen by the player in the player's rest frame:
\al{
\fv\Xi
	&=	L\fn{\bs u}\paren{\fv\xi-\fv{x_\P}}
	\approx
		L\fn{\bs u}\fv\xi.
			\label{Xi from xi} 
}
We parametrize it as
\al{
\fv\Xi
	&=	\bmat{-R\\ R\sin\Theta\cos\Phi\\ R\sin\Theta\sin\Phi\\ R\cos\Theta},
		\label{Xi from Theta Phi}
}
where $R$ is very large.\footnote{
Be careful with the abuse of notation: This $R$ has nothing to do with the rotation matrix in Eq.~\eqref{rotation matrix}.
}
How is $\paren{\Theta,\Phi}$ related to $\paren{\theta,\phi}$, used to pick up the drawn texture point $\paren{S,T}$ in Eq.~\eqref{ST vs theta phi}?

Equation~\eqref{Xi from xi} implies that
\al{
\fv\xi
	&=
		L\fn{-\bs u}\fv\Xi.
			\label{xi from Xi}
}
For a given direction $\paren{\Theta,\Phi}$, we determine $\fv\Xi$ by Eq.~\eqref{Xi from Theta Phi} up to the overall constant $R$, and then obtain $\fv\xi$ by Eq.~\eqref{xi from Xi}.
Once $\fv\xi$ is known, $\paren{\theta,\phi}$ are given by Eq.~\eqref{theta phi}, and then the corresponding texture coordinates $\paren{S,T}$ to be drawn are obtained by Eq.~\eqref{ST vs theta phi}. Note that the overall normalization $R$ drops out of the final expression since the coordinates $\fv\Xi$ always appear as ratios in Eq.~\eqref{theta phi}. In practice, we may put $R=1$.

\subsection{2D background}

2D games can be handled similarly.
We explain how to draw a background using orthogonal projection in the $3=2+1$D spacetime.
We assume for simplicity that the player's view does not rotate with respect to the background.
Suppose that the background is at rest at the reference frame $\Set{x}$, where $x=\paren{x^0,\bs x}=\paren{x^0,x^1,x^2}$ is a 3($2+1$)D spacetime vector, whose spatial components consist of a spatial vector $\bs x=\paren{x^1,x^2}$.\footnote{
Recall that $x^2$ does not denote $x$-squared but the second component of $\bs x=\paren{x^1,x^2}$ throughout this paper, unless otherwise stated.
}

Let $\paren{S,T}$ be the texture coordinates of the background, with $0\leq S\leq 1$ and $0\leq T\leq 1$.
Suppose that the background texture is mapped onto a region $x^1_\tx{min}\leq x^1\leq x^1_\tx{max}$ and $x^2_\tx{min}\leq x^2\leq x^2_\tx{max}$:
\al{
S	&=	{x^1-x_\tx{min}^1\over x^1_\tx{max}-x^1_\tx{min}},	&
T	&=	{x^2-x_\tx{min}^2\over x^2_\tx{max}-x^2_\tx{min}}.
	\label{ST coordinate}
}
When the player's velocity is $\bs u_\tx{P}$ in the reference frame, we want to know which point in the texture coordinate $\paren{S,T}$ is picked up for a given point in the player's rest frame $\Set{X}$:
\al{
X	&=	L\fn{\bs\uP}\paren{x-\xP},
}
where we have chosen the origin of $X$ to be the point corresponding to $\xP$ in the reference frame.

The player sees the world sliced by its PLC.
Therefore, a natural way to draw the 2D world is to project the player's PLC onto the $X^1$-$X^2$ plane.
Let the drawn region on screen be $X_\tx{min}^1\leq X^1\leq X_\tx{max}^1$ and $X_\tx{min}^2\leq X^2\leq X_\tx{max}^2$.
A spatial point $\bs X=\paren{X^1,X^2}$ in this region has the time coordinate
\al{
X^0=-\ab{\bs X}=-\sqrt{\paren{X^1}^2+\paren{X^2}^2}
}
on the player's PLC.
The corresponding point in the reference frame is
\al{
x
	&=	\xP+L\fn{-\bs\uP}\bmat{-\sqrt{\paren{X^1}^2+\paren{X^2}^2}\\ X^1\\ X^2}.
}
From this, we can read off the values of the spatial components $\paren{x^1,x^2}$, which can be put in Eq.~\eqref{ST coordinate} to get the corresponding texture coordinate $\paren{S,T}$.

By this PLC formalism, one may improve the drawing of the world by an equal-time slice in the player's rest frame, employed, e.g., in Ref.~\cite{VelRap}.

\subsection{Approximately rigid body}\label{rigid body section}
In the relativistic world, it is impossible to have an exactly rigid body, as it violates the locality.
However, a thing is usually introduced as a rigid body, described by a set of polygons, in computer games.
We show how to treat such a thing approximately.

Suppose that a body $B$ has $N$ vertices that are specified by spatial $d$ vectors
\al{
\bs\xi_a
	&=	\paren{\xi_a^1,\dots,\xi_a^d}
	=	\bmat{\xi_a^1\\ \vdots\\ \xi_a^d},
		\label{polygon vertices}
}
where $a=1,\dots,N$.
Each $d$ vector $\bs\xi_a$ specifies the position of the corresponding vertex measured from $B$'s reference point, somewhere near its center. Hereafter, we call this point, somewhat sloppily, $B$'s center.

Let $R_\ttt$ be a $d\times d$ rotation matrix that represents the orientation of $B$, which differs at each spacetime point $\fv{x_\ttt}$ that consists of the worldline $\Set{\fv{x_1},\fv{x_2},\dots}$.\footnote{
In practice, one can also store $R_\ttt$ along with each $\fv{x_\ttt}$ in the worldline data of $B$.
}
At each worldline point $\fv{x_\ttt}$, the object $B$ has the velocity~\eqref{eq:u_j} that is uniform for all the polygon vertices~\eqref{polygon vertices}: $B$'s position in its rest frame is given by
\al{
\fv{X_\ttt}=L\fn{\bs u_\ttt}\fv{x_\ttt}.
}
More precisely, $\fv{X_\ttt}$ represents the position of $B$'s center in its rest frame, at the \ttt th point in $B$'s worldline.

We approximate that $B$ is a rigid body, i.e., $R_\ttt$ rotates all the vertices~\eqref{polygon vertices} simultaneously: The position of each vertex, say the $a$th one, in $B$'s rest frame is given by
\al{
\bs X_{a,\ttt}
	&=	\bs X_\ttt+R_\ttt\bs\xi_a,
		\label{Xajspace}
}
when $B$ is at the $j$th point on its worldline.
How can one determine its time coordinate~$X_{a,\ttt}^0$?
We approximate that each worldline of the vertex is parallel in spacetime to that of $B$'s center:
It is proportional to $\fv{X_{\ttt+1}}-\fv{X_\ttt}$.
Then we can obtain the intersection between the worldline of the $a$th vertex and the player's PLC from the condition
\al{
\lip{\fv{X_{a,\ttt}}-\fv{X_\P}}^2
	&=	0,
}
i.e.,
\al{
X_{a,\ttt}^0
	&=	X_\P^0-\ab{\bs X_{a,\ttt}-\bs X_\P},
		\label{Xaj0}
}
where $X_\P$ is the player's position in $B$'s rest frame.
In the original reference frame, the position of the $a$th vertex is
\al{
\fv{x_{a,\ttt}}
	&=	L\fn{-\bs u_\ttt}\fv{X_{a,\ttt}},
}
where the time and spatial components of $\fv{X_{a,\ttt}}$ are given by Eqs.~\eqref{Xajspace} and \eqref{Xaj0}, respectively.
This $x_{a,\ttt}$ can be transformed to that in the player's rest frame as in Sec.~\ref{formulation} when the world is drawn.

Rotation of the rigid body, or, more explicitly, the information on $R_\ttt$ above, can be most easily handled by using quaternions; see Appendix~\ref{quaternion} for a review.
We specify the rotation of a rigid body from its basic position by an angle $\theta$ around a direction $\bh\ell$ using the quaternion
\al{
\Theta\fn{\bh\ell,\theta}
	&:=	\cos{\theta\over2}+\sin{\theta\over2}\,\bh\ell\cdot\bs\iq,
}
where $0\leq\theta\leq\pi$ and $\ab{\bh\ell}=1$.
There are three independent degrees of freedom in total in $\theta$ and $\bh\ell$, which coincide with the physical degrees of freedom, namely, a unit vector to specify the front (upward, or whatever) direction, and an angle to determine the rotation around that direction.

2D games can be handled similarly.
Usually an object is drawn by a rectangular picture.
We may divide the picture into $N\times N'$ cells, where $N$ and $N'$ are appropriate numbers, typically of order 10 to 100.
Then we can treat the $NN'$ vertices as above, and map the texture on each cell.

\subsection{Doppler effect}\label{Doppler section}
Light that comes into our eyes has a spectrum; namely, its intensity $I$ is a function of the angular frequency $\omega$:
\al{
I\fn{\omega},
	\label{spectrum}
}
where $\omega$ is related to the frequency $f$ and the wavelength $\lambda$ by
\al{
\omega
	&=	2\pi f
	=	{2\pi c\over\lambda}.
}
Here and hereafter, we recover $c$ from the natural units in the expressions involving $\omega$.
When light ray points in a direction $\bh n$, where $\ab{\bh n}=1$ is a unit $d$ vector, it is known that the wave $D$ vector\footnote{
In an nonrelativistic context, the $d$ vector $\bs k/c=\paren{\omega/c}\bh n$ is usually called the wave vector.
}
\al{
\fv k
	&:=	\omega\bmat{1\\ \bh n}
		\label{wave vector}
}
is proportional to the photon's $D$ momentum and that $\fv k$ transforms as a $D$ vector under the Lorentz transformation; see Appendix~\ref{energy momentum}.

Suppose that a source $S$ and an observer $O$ have their velocities $\fv{u_S}$ and $\fv{\uO}$, respectively, in the reference frame.
Let us consider light of an angular frequency $\Omega$ pointing towards $\bh N$ with $\ab{\bh N}=1$ in $S$'s rest frame:
\al{
\fv K
	&=	\Omega\bmat{1\\ \bh N}.
		\label{K vector}
}
The direction $\bh N$ can be obtained as follows: When $O$ at $\fv{\xO}$ observes the light emitted from $\fv{x_S}$, then $\fv{x_S}$ must be on $\PLC\fn{\fv{\xO}}$:
\al{
\lip{\fv{\xO}-\fv{x_S}}^2=0
}
and $\xO^0>x_S^0$. In this case, $\fv K$ is proportional to
\al{
\fv{X_{OS}}
	:=	L\fn{\bs u_S}\paren{\fv{\xO}-\fv{x_S}}.
}
Then we obtain
\al{
\bh N
	&=	{\bs X_{OS}\over X_{OS}^0}.
}

The light~\eqref{K vector} is observed by $O$ as the wave vector
\al{
\fv{K'}
	&:=	L\fn{\bs \uO}L\fn{-\bs u_S}\fv K.
}
The angular frequency $\Omega'$ observed by $O$ is then the time component of $\fv{K'}$, namely $K^{\pr 0}$.
Note that $\Omega'$ is proportional to $\Omega$:
\al{
\Omega'=C\Omega,
}
where $C$ is a constant that can be computed from the above procedure.
Then the spectrum~\eqref{spectrum} is perceived by $O$ as a new spectrum $I_\tx{O}$:
\al{
I_\tx{O}\fn{\Omega}
	&=	I\fn{\Omega/C}.
} 
This is all needed to determine the Doppler effect on light. 

Most fundamentally, one should consult the above procedure.
In practice, it does not work. Why?
In the RGB color model used in computers, a color is specified by three numbers $\paren{R,G,B}$ that roughly represent the perception of three types of cone cells in our eyes.
For a given spectrum~$I\fn{\omega}$, the values of $R$, $G$, $B$ are obtained, up to an overall normalization of $\paren{R,G,B}$, by
\al{
R	&=	\int_{\lambda_\tx{min}}^{\lambda_\tx{max}}\df\lambda\,\ol r\fn{\lambda}I\fn{2\pi c\over\lambda},\\
G	&=	\int_{\lambda_\tx{min}}^{\lambda_\tx{max}}\df\lambda\,\ol g\fn{\lambda}I\fn{2\pi c\over\lambda},\\
B	&=	\int_{\lambda_\tx{min}}^{\lambda_\tx{max}}\df\lambda\,\ol b\fn{\lambda}I\fn{2\pi c\over\lambda},
}
where the CIE standard observer color matching functions $\ol r$, $\ol g$, and $\ol b$ are given in Ref.~\cite{Wikipedia_color} and $\lambda_\tx{min}=380\,\tx{nm}$ and $\lambda_\tx{max}=780\,\tx{nm}$ in the CIE 1931 color space.
If $I\fn{\omega}$ were given to specify a color, instead of $\paren{R,G,B}$, then we could calculate the new values of $\paren{R',G',B'}$ for $O$:
\al{
R'	&=	\int_{\lambda_\tx{min}}^{\lambda_\tx{max}}\df\lambda\,\ol r\fn{\lambda}I\fn{2\pi c\over\lambda C},\\
G'	&=	\int_{\lambda_\tx{min}}^{\lambda_\tx{max}}\df\lambda\,\ol g\fn{\lambda}I\fn{2\pi c\over\lambda C},\\
B'	&=	\int_{\lambda_\tx{min}}^{\lambda_\tx{max}}\df\lambda\,\ol b\fn{\lambda}I\fn{2\pi c\over\lambda C}.
}
The problem is that one needs the full spectrum $I\fn{\omega}$ for the exact relativistic computation, whereas one cannot know the functional form of $I\fn{\omega}$ from merely the three numbers $R$, $G$, and $B$.

Here we present a naive procedure to mimic the Doppler effect.
For blackbody radiation, $\paren{R,G,B}$ is known as a function of the temperature $T$. That is, we know the functions
\al{
R\fn{T}, \quad G\fn{T}, \quad B\fn{T}
	\label{RGB}
}
for the blackbody radiation; see, e.g., Ref.~\cite{blackbody} for possible approximate fit functions.\footnote{\label{approximate color fit}
The explicit form in Ref.~\cite{blackbody} is
\als{
R\fn{T}
	&=	\begin{cases}
			255	&	T<6688\,\tx{K},\\
			329.70\paren{T-6000\,\tx{K}\over 100\,\tx{K}}^{-0.133205}
				&	T>6688\,\tx{K},
		\end{cases}
		\\
G\fn{T}
	&=	\begin{cases}
			0	&	T<505\,\tx{K},\\
			99.47\ln{T\over100\,\tx{K}}-161.12
				&	505\,\tx{K}<T<6503\,\tx{K},\\
			287.12\paren{T-6000\,\tx{K}\over100\,\tx{K}}^{-0.0755148}	&	T>6503\,\tx{K},
		\end{cases}
		\\
B\fn{T}
	&=	\begin{cases}
			0	&	T<1904\,\tx{K},\\
			138.52\ln{T-1000\,\tx{K}\over100\,\tx{K}}-305.04
				&	1904\,\tx{K}<T<6700\,\tx{K},\\
			255	&	T>6700\,\tx{K}.
		\end{cases}
}
}
For a given set of $\paren{R,G,B}$, we may then naively compute color-specific temperatures $\paren{T_R,T_G,T_B}$ by inverting Eq.~\eqref{RGB}.\footnote{
A color function, say, $T_G$ may not be a single-valued function of $G$ but a twofold one, as in footnote~\ref{approximate color fit}.
Even then, one may in practice take the lower (higher) value of $T_G$ when $R>B$ ($R<B$) for a given set $\paren{R,G,B}$.
}
Physically, a temperature roughly behaves as an energy, which is the 0th component of the energy-momentum $D$ vector, proportional to the wave vector~\eqref{wave vector} for photons. Therefore, the new color temperature for $O$ would be estimated by $T/C$ for each color. Then one may get the naive Doppler-shifted colors as
\al{
R\fn{T_R/C}, \quad G\fn{T_G/C}, \quad B\fn{T_B/C}.
}

We have shown how to mimic the Doppler effect within the RGB color scheme.
As said, the full treatment requires the inclusion of the spectrum of the light for each pixel, instead of the current approximation using the three RGB numbers. That will allow visualization of the currently invisible ultraviolet and infrared lights when one is boosted significantly by a Lorentz transformation.

\section{Summary}\label{summary}
We have shown how to implement special relativity in computer games. The past light cone is used for the foliation of the world without violating causality. This is the first realization of interacting nonplayer characters that perceive the world and react to others, both relativistically without violating causality, under fully relativistic time evolution. In the formulation, the notion of the Lorentz-covariant velocity is extensively used, instead of the more widely used noncovariant velocity. We have shown several ideas to approximate the relativistic world such as the relativistic generalization of the skydome, rigid body, and Doppler effect.

You may find an implementation of the ideas presented in this paper in Ref.~\cite{sogebu}.

\subsection*{Acknowledgements}
The idea of a relativistic game occurred to K.O.\ while watching Ref.~\cite{Haruhi}.
K.O.\ thanks all the participants of the class ``Making relativistic games'' at Osaka University during 2009--2012 and 2015.
The work of K.O.\ is partially supported by JSPS KAKENHI Grant Nos.~15K05053 and 23104009.
\appendix
\section*{Appendix}

\section{Relativistic energy and momentum}\label{energy momentum}
From the Lorentz-covariant velocity $\fv u$, we may define the Lorentz-covariant $D$ momentum
\al{
\fv p	&=	m\fv u,
}
where $m$ is the mass of the particle.
Alternatively, we may give the $D$ momentum $\fv p$ first, and define $m$ to be its Lorentz-invariant norm
\al{
m	&:=	\sqrt{-\lip{\fv p}^2},
		\label{mass defined}
}
which is a fixed constant for each species of particle.\footnote{
We have assumed that $\fv p$ is either time-like $\lip{\fv p}^2<0$ or light-like $\lip{\fv p}^2=0$.
A particle having space-like momentum $\lip{\fv p}^2>0$ moves faster than the speed of light, and is the tachyon discussed in footnote~\ref{tachyon}.
}
The $D$ velocity is then given as $\fv u=\fv p/m$ for a massive particle.

The time component of $\fv p$ is identified as the energy of the particle:
\al{
E	&=	p^0.
}
From the definition of mass~\eqref{mass defined}, we obtain
\al{
E	&=	\sqrt{m^2+\bs p^2}.
}
When the particle is at rest $\bs p=0$, its energy reduces to its mass, and we get Einstein's celebrated formula for the rest energy
\al{
E	&=	mc^2,
}
where we have tentatively recovered the speed of light $c$.
Note that the ratio of momentum to energy gives the ordinary noncovariant velocity
\al{
{\bs p\over E}
	&=	{\bs u\over u^0}
	=	{\bs u\over \gamma\fn{\bs u}}
	=	\bs v
}
in any coordinate system.

For a massless particle such as a photon, of which light consists, the velocity $\fv u$ cannot be defined, and rather its momentum $\fv p$ is a suitable physical quantity.
In quantum mechanics, photon's $D$ momentum $\fv p$ is related to the \emph{wave $D$ vector} $\fv k$ by
\al{
\fv p
	&=	\hbar\fv k,
	\label{wave vector}
}
where $\hbar\simeq10^{-34}\,\tx{J}\tx{s}$ is the \emph{reduced Planck constant} and, tentatively recovering $c$ from the natural units,
\al{
\fv p
	&=	\bmat{E/c\\ \bs p},&
\fv k
	&=	\bmat{\omega/c\\ \bs k},
}
in which $\omega$ and $\bs k$ are the \emph{angular velocity} and the \emph{wave vector}, respectively.
Note that
\al{
\omega
	&=	2\pi f,	&
k^i
	&=	{2\pi\over \lambda^i}&
	&(i=1,\dots,d),
}
where $f$ is the frequency and $\lambda^i$ is the wavelength for the $i$th direction.
Note also that the photon momentum $D$ vector is light-like:
\al{
\lip{\fv p}^2
	=	\lip{\fv k}^2
	&=	0.
}
The relation~\eqref{wave vector} is used when taking into account the Doppler effect in Sec.~\ref{Doppler section}.

\section{Lorentz contraction and dilation}\label{Lorentz contraction and dilation}
We show that a measuring rod moving towards the observer looks \emph{longer} than when it is at rest, because of the Lorentz ``contraction'', due to the PLC foliation.
For simplicity, we work in the $D=1+1$-dimensional spacetime with only $d=1$ spatial dimension.

We consider a measuring rod that has a length $\ell$ in its rest frame.
Suppose it stays at rest forever in its rest frame. Then its ``worldsheet'' can be written in the rest frame as
\al{
W_\tx{rod}
	&=	\Set{\fv X=\paren{X^0,X^1}|-\infty<X^0<\infty, \quad 0\leq X^1\leq\ell}.
}
We call $X^1=0$ and $\ell$ the left and right ends of the rod, respectively.

\begin{figure}[tp]
\begin{center}
\includegraphics[width=0.4\textwidth]{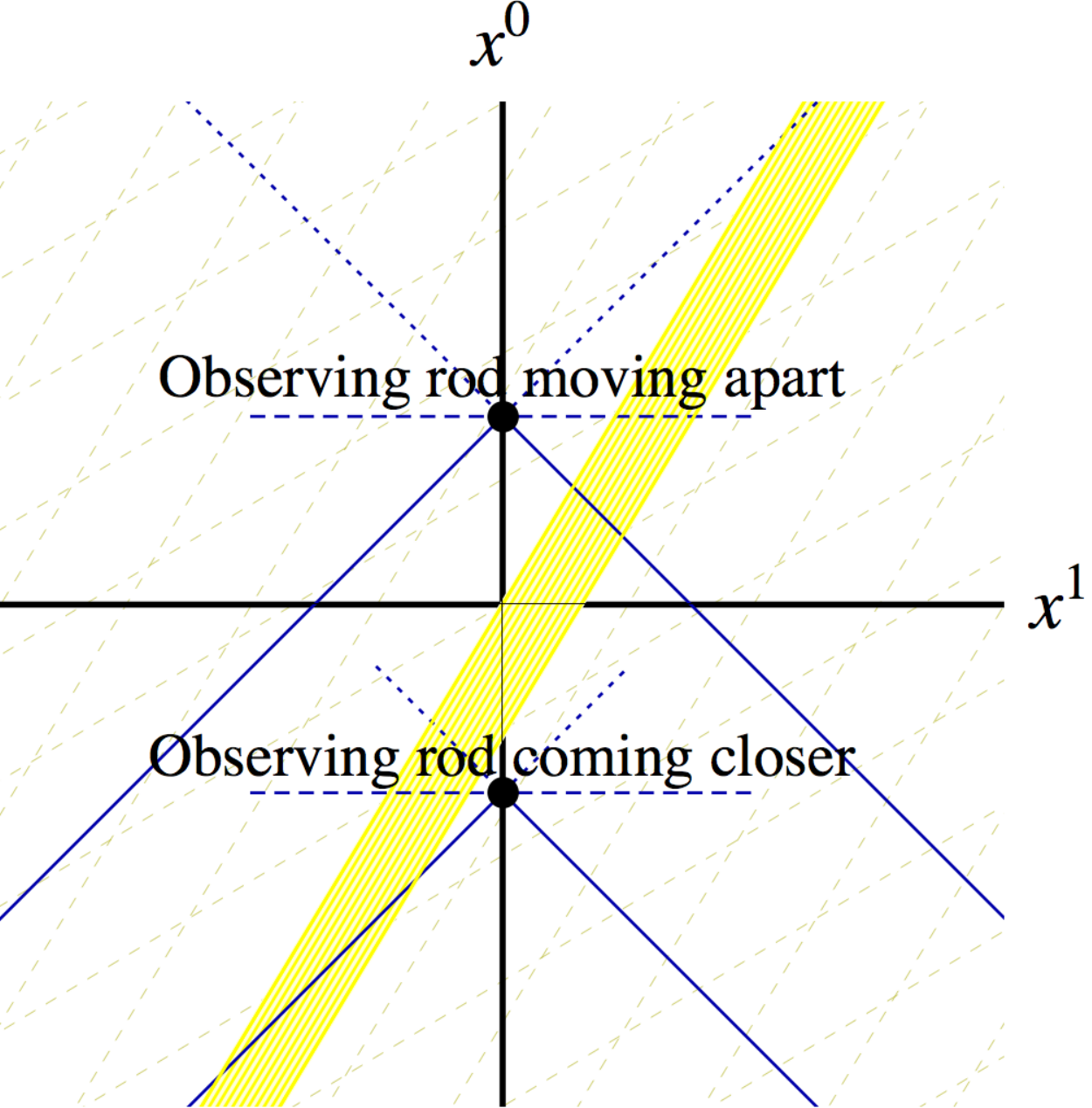}
\caption{Schematic figure in the reference frame $\paren{x^0,x^1}$ in which the observer stays at rest at $x^1=0$.
The worldsheet of the measuring rod is collectively represented by the yellow band. Solid and dotted diagonal blue lines are the PLC and FLC of the observer, respectively. Thin dashed dark-yellow lines represent the rod's rest frame. Blue dashed horizontal lines are equal-time slices for the observer.}\label{Lorentz contraction schematic}
\end{center}
\end{figure}

Suppose that the rod has a velocity $\fv u=\paren{\gamma\fn{u^1},\,u^1}$ in the reference frame, where the Lorentz factor is
\al{
u^0=\gamma\fn{u^1}=\sqrt{1+\paren{u^1}^2}\geq1.
}
Here and hereafter in this section, we do not make the unnecessary distinction between $\bs u$ and $u^1$ as we work in the $d=1$ spatial dimension.
See Fig.~\ref{Lorentz contraction schematic} for a schematic plot.
The reference frame $\fv x$ is obtained from the rod's rest frame $\fv X$ by
\al{
\fv x=L\fn{-u^1}\fv X
}
where
\al{
L\fn{-u^1}
	&=	\bmat{\gamma\fn{u^1}&u^1\\ u^1&\gamma\fn{u^1}}.
}
We see that the time coordinate of the rest frame is written as $X^0=\gamma\fn{u^1}x^0-u^1\,x^1$ and hence
\al{
x^1	&=	v\,x^0+{1\ov\gamma\fn{u^1}}X^1,
}
where $v:=u^1/\gamma\fn{u^1}$ is the noncovariant velocity.
In the reference frame $\Set{\fv x}$, the rod appears to move with the (noncovariant) velocity $v$, and its length measured in $x^1$ is \emph{contracted} to be $\ell/\gamma\fn{u^1}$ as $X^1$ varies from 0 to $\ell$.
That is, the worldsheet of the rod can be written in the reference frame as
\al{
W_\tx{rod}
	&=	\Set{\fv x=\paren{x^0,v\,x^0+\delta x^1}|-\infty<x^0<\infty,\quad0\leq\delta x^1\leq{\ell\ov\gamma\fn{u^1}}},
}
where $\delta x^1:=X^1/\gamma\fn{u^1}$.
This is the celebrated \emph{Lorentz contraction}.

However, the contracted length $\ell/\gamma\fn{u^1}$ is the length at an \emph{equal-time slice} $x^0=\tx{const.}$, which is not really a length measured on any observer's PLC.
The equal-time slices are represented by the blue dashed horizontal lines in Fig~\ref{Lorentz contraction schematic}.
To see the real measured length, let us prepare an observer staying at rest at $x^1=0$ in the reference frame.
We restrict ourselves to the case of a right-moving rod $u^1>0$ without loss of generality.
The observer sees the rod from $\fv{\xO}=\paren{\xO^0,0}$.

When the rod is moving away from the observer, $\xO^0>0$,
the left and right ends of the rod intersect with $\PLC\fn{\fv{\xO}}$ at $\fv{x_\tx{L}}$ and $\fv{x_\tx{R}}$:
\al{
\fv{x_\tx{L}}
	&=	\paren{\gamma\fn{u^1}-u^1}\bmat{\gamma\fn{u^1}\xO^0\\ u^1\,\xO^0},	&
\fv{x_\tx{R}}
	&=	\paren{\gamma\fn{u^1}-u^1}\bmat{\gamma\fn{u^1}\xO^0-\ell\\ u^1\,\xO^0+\ell}.
}
Therefore, the observed length of the rod is
\al{
\ell_\tx{obs}
	&=	x_\tx{R}^1-x_\tx{L}^1
	=	\paren{\gamma\fn{u^1}-u^1}\ell.
		\label{departing rod}
}
Using the identity $\gamma\fn{u}-u=\paren{\gamma\fn{u}+u}^{-1}$, we see that this is \emph{shorter} than the Lorentz contraction $\ell/\gamma\fn{u^1}$.

When the rod is moving towards the observer, $\xO^0<-\ell/u^1$,
the left and right ends of the rod intersect with $\PLC\fn{\fv{\xO}}$ at $\fv{x_\tx{L}}$ and $\fv{x_\tx{R}}$:
\al{
\fv{x_\tx{L}}
	&=	{1\ov\gamma\fn{u^1}-u^1}\bmat{\gamma\fn{u^1}\xO^0\\ u^1\,\xO^0},	&
\fv{x_\tx{R}}
	&=	{1\ov\gamma\fn{u^1}-u^1}\bmat{\gamma\fn{u^1}\xO^0+\ell\\ u^1\,\xO^0+\ell}.
}
The observed length of the rod is
\al{
\ell_\tx{obs}
	&=	x_\tx{R}^1-x_\tx{L}^1
	=	\paren{\gamma\fn{u^1}+u^1}\ell,
		\label{coming rod}
}
where we have used the identity $1/\paren{\gamma\fn{u}-u}=\gamma\fn{u}+u$. We see that the observed length is \emph{longer} than $\ell$ when the rod is moving towards the observer.

\begin{figure}[tp]
\begin{center}
\hfill
\includegraphics[width=0.4\textwidth]{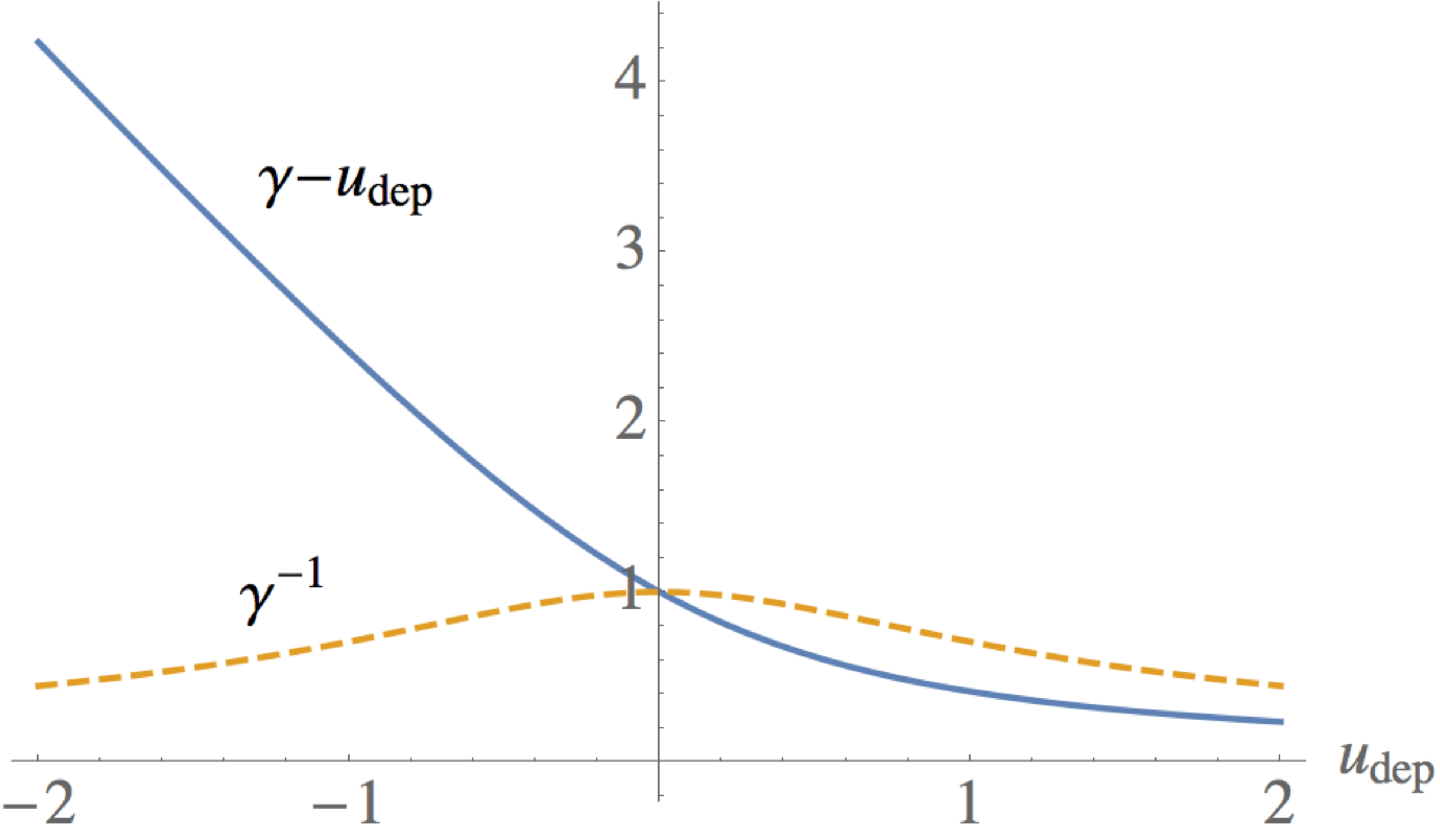}\hfill
\includegraphics[width=0.4\textwidth]{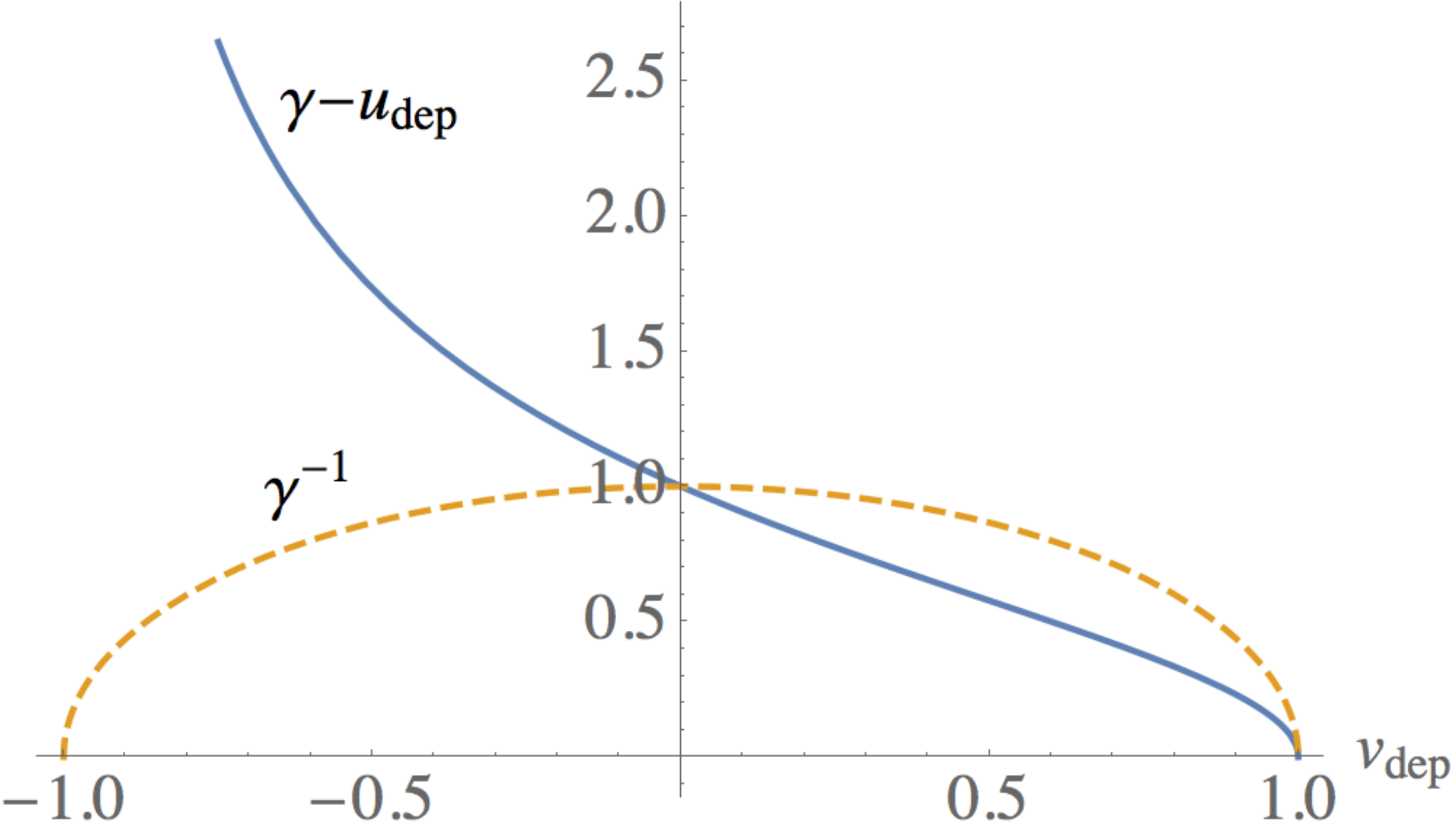}\hfill\mbox{}
\caption{The blue solid line shows the observed Lorentz factor $\gamma\fn{u_\tx{dep}}-u_\tx{dep}$ as a function of $u_\tx{dep}$ (left) and $v_\tx{dep}$ (right). For comparison, the conventional Lorentz factor $1/\gamma\fn{u_\tx{dep}}$ is also plotted in the yellow dashed lines.}\label{Lorentz factor figure}
\end{center}
\end{figure}

To summarize, when we write the departing velocity of the rod $u_\tx{dep}$, which is $u^1$ and $-u^1$ in Eqs.~\eqref{departing rod} and \eqref{coming rod}, respectively, the apparent length of the rod is
\al{
\ell_\tx{obs}
	&=	\paren{\gamma\fn{u_\tx{dep}}-u_\tx{dep}}\ell.
}
In the left panel of Fig.~\ref{Lorentz factor figure}, we show this factor as a function of $u_\tx{dep}$. We also show it in the right as a function of the noncovariant departing velocity $v_\tx{dep}:=u_\tx{dep}/\gamma\fn{u_\tx{dep}}$.

\section{Evolution in co-moving frames}\label{co-moving evolution}
To cultivate physical intuition, it is instructive to formulate how the player, in its rest frame at each moment, sees the motion of the others that are always on the player's PLC.
The formulation in Sec.~\ref{others' time evolution} is better than the one in this section in the sense that it takes into account the time dilation more accurately in the discretized time steps, while the formulation in this section makes the time evolution more tractable in an analytic treatment.

Suppose that the player's position and velocity at its proper time $\tau_\P$ are $\fv\xP$ and $\fv\uP$, respectively, in the reference frame.
After an infinitesimal time $\Delta\tau_\P$ felt by the player, we shift the player's proper time, position, and velocity as\footnote{
$\lip{\fv{u_\P'}}^2=-1$ holds up to the $\Or{\paren{\Delta\tau_\P}^2}$ term because $\fv{u_\P}\cdot\fv{a_\P}=0$. Throughout this section, we shall neglect the quadratic and higher-order terms of infinitesimal quantities. In an actual implementation with finite time step, one may always reset any time component of the velocity $u^{\pr0}$ to $\sqrt{1+\bs u^{\pr2}}$ whenever necessary.
}
\al{
\tau_\P
	&\longmapsto
		\tau_\P'=\tau_\P+\Delta\tau_\P,\nn
\fv\xP
	&\longmapsto
		\fv{\xP'}=\fv\xP+\fv\uP\,\Delta\tau_\P,\nn
\fv\uP
	&\longmapsto
		\fv{\uP'}=\fv\uP+\fv{a_\P}\,\Delta\tau_\P,
			\label{player's co-moving evolution}
}
where $\fv{a_\P}$ is the player's acceleration at the spacetime point $\fv\xP$ in the reference frame.

In the PLC foliation, all the other objects are located on $\PLC\fn{\fv\xP}$:
The $n$th-particle's position $\fv{x_n}$ satisfies $x_n^0<\xP^0$ and
\al{
\lip{\fv{x_n}-\fv\xP}^2
	&=	0.
		\label{PLC condition for n and P}
}
In the iteration~\eqref{player's co-moving evolution}, the $n$th particle moves onto
\al{
\fv{x_n}
	&\longmapsto	\fv{x_n'}=\fv{x_n}+\fv{u_n}\,\Delta\tau_n,\\
\fv{u_n}
	&\longmapsto	\fv{u_n'}=\fv{u_n}+\fv{a_n}\,\Delta\tau_n,
}
where $\fv{a_n}$ is the acceleration of the $n$th particle at $\fv{x_n}$ in the reference frame.

As the player has moved to $\fv{x_\P'}$, the hypersurface seen by the player in the next step is $\PLC\fn{\fv{x_\P'}}$.
The amount of the proper-time difference $\Delta\tau_n$ is determined by the requirement that $\fv{x_n'}$ is on $\PLC\fn{\fv{x_\P'}}$:
\al{
\lip{\fv{x_n'}-\fv{x_\P'}}^2=0
}
with $x_n^{\pr0}<x_\P^{\pr0}$. Concretely, we obtain
\al{
\Delta\tau_n
	&=	\lip{\fv{x_n}-\fv{x_\P'},\fv{u_n}}-\sqrt{\lip{\fv{x_n}-\fv{x_\P'},\fv{u_n}}^2+\lip{\fv{x_n}-\fv{x_\P'}}^2}\nn
	&=	\lip{\fv{x_n}-\fv{x_\P'},\fv{u_n}}-\sqrt{\lip{\fv{x_n}-\fv{x_\P'},\fv{u_n}}^2
			-2\lip{\fv{x_n}-\fv{x_\P},\fv{u_\P}}\,\Delta\tau_\P
			-\paren{\Delta\tau_\P}^2
			},
			\label{delta tau n}
}
where we have used $\lip{\fv{u_n}}^2=-1$ and Eq.~\eqref{PLC condition for n and P}, as well as $\lip{\fv{u_\P}}^2=-1$, in the first and second steps, respectively.
Note that $\lip{\fv{x_n}-\fv{x_\P'},\fv{u_n}}>0$.
In the limit $\Delta\tau_\P\to0$, we get
\al{
\Delta\tau_n
	&\approx
		{\lip{\fv{x_n}-\fv{x_\P},\fv{u_\P}}\ov\lip{\fv{x_n}-\fv{x_\P},\fv{u_n}}}\,\Delta\tau_\P,
				\label{delta tau n limit}
}
where $\approx$ indicates that the quadratic and higher-order terms of the infinitesimal quantities are neglected.\footnote{
In an actual implementation with finite $\Delta\tau_\P$, Eq.~\eqref{delta tau n} is more usable than Eq.~\eqref{delta tau n limit}.
}

As in footnote~\ref{central frame footnote}, let us introduce the central-frame coordinates $\fv X$ in which the player at its proper time $\tau_\P$ looks at rest at the spacetime origin: $\fv{X_\P}\fn{\tau_\P}=0$ and $\fv{U_\P}\fn{\tau_\P}=0$.
The corresponding central-frame position vector $\fv X$ to the reference-frame one $\fv x$ is
\al{
\fv X	&=	L\fn{\bs \uP\fn{\tau_\P}}\paren{\fv x-\fv\xP\fn{\tau_\P}}.
	\label{X from x}
}
The central frame differs at each proper time of the player $\tau_\P$, and we call it the $\mc C\fn{\tau_\P}$ frame; we also call them collectively the co-moving frames.
In particular, the player is always at rest at the origin in the co-moving frames: $\fv{\mc X_\P}=\paren{0,\bs 0}$ and $\fv{\mc U_\P}=\paren{1,\bs 0}$.
Here and hereafter, we write the quantities in the co-moving frames in a curly upper-case letter. We continue to write the quantities in the reference frame and in each rest frame (at a particular moment) in lower and upper cases, respectively.
We consider the foliation by the player's PLCs and formulate the time evolution as always pulled back to the player's central frame at each moment.

At $\tau_\P$, the co-moving-frame coordinates $\fv{\mc X}$ are identified with the $\mc C\fn{\tau_\P}$-frame coordinates $\fv X$:
\al{
\left.\fv{\mc X}\right|_{\tau_\P}
	&=	\fv X,	\label{identification at first}
}
where $\fv X$ is given in Eq.~\eqref{X from x}.
For a given acceleration $\fv{A_\P}=\paren{0, \bs A_\P}$ in the $\mc C\fn{\tau_\P}$ frame, the player's next location and the velocity after an infinitesimal time $\Delta \tau_\P$, felt by the player, are\footnote{
The reference-frame acceleration reads $\bs a_\P=\bs A_\P+\paren{\gamma\fn{\bs u_\P}-1}\paren{\bh u_\P\cdot\bs A_\P}\bh u_\P$, from which we may obtain the reference-frame velocity in the next step: $\bs u_\P'=\bs u_\P+\bs a_\P\Delta\tau_\P$.
}
\al{
\fv{X_\P}=\bmat{0\\ \bs 0}
	&\longmapsto
		\fv{X_\P'}
		:=	\bmat{0\\ \bs 0}+\bmat{1\\ \bs 0}\Delta \tau_\P
		=	\bmat{\Delta\tau_\P\\ \bs 0},\\
\fv{U_\P}=\bmat{1\\ \bs 0}
	&\longmapsto
		\fv{U_\P'}
		:=	\bmat{1\\ \bs 0}+\bmat{0\\ \bs A_\P}\Delta \tau_\P
		=	\bmat{1\\ \bs A_\P\Delta\tau_\P}.
}

\begin{figure}[tp]
\begin{center}
\includegraphics[width=0.7\textwidth]{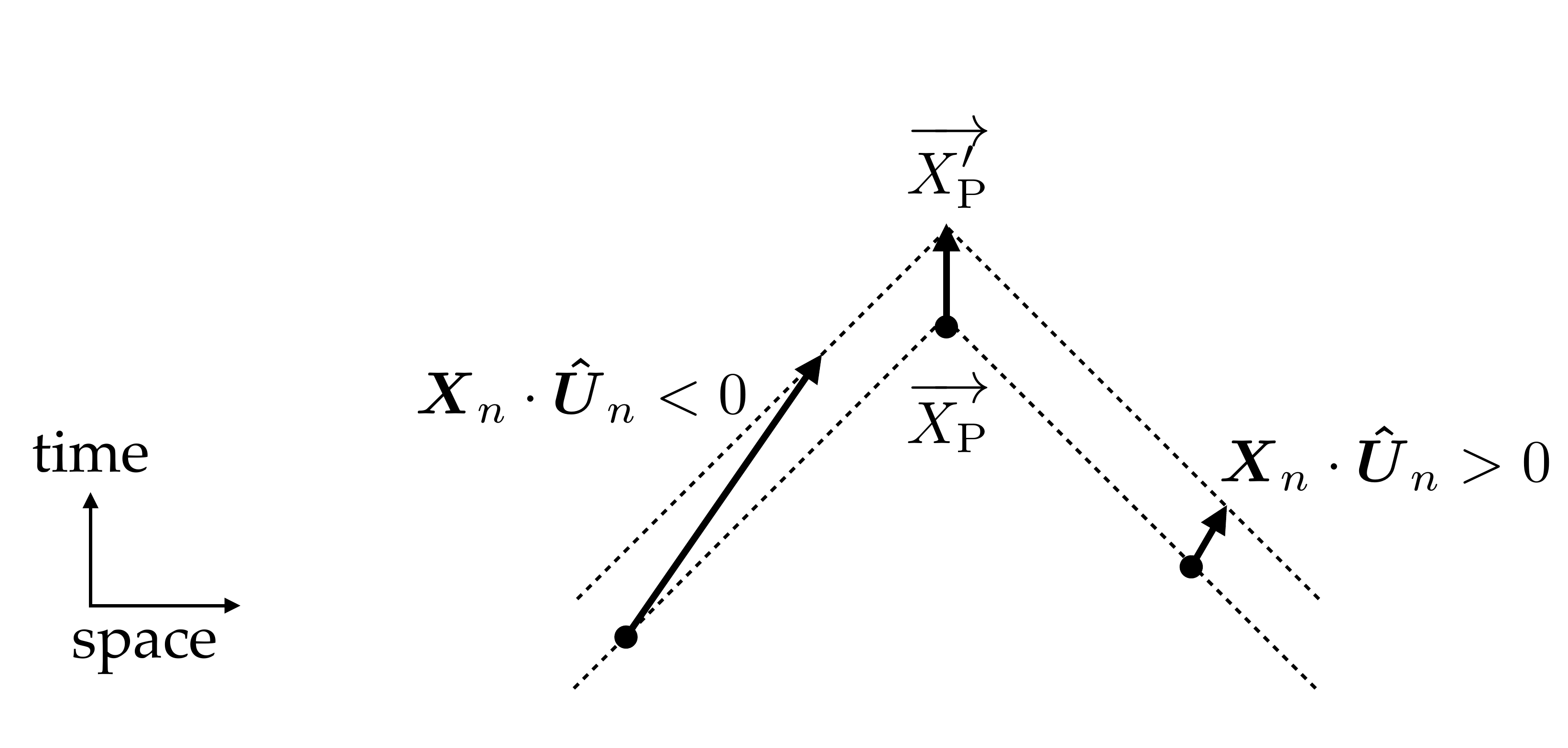}
\caption{Schematic plot to show the difference between the situations where the $n$th particle is coming closer to the player, $\bs X_n\cdot\bh U_n<0$ in the player's central frame, and going away from the player, $\bs X_n\cdot\bh U_n>0$.}\label{delta tau n figure}
\end{center}
\end{figure}

At $\tau_\P$, every object is located on $\PLC\fn{\fv\xP}$. In the $\mc C\fn{\tau_\P}$ frame, we may parametrize the location of the $n$th particle as 
\al{
\left.\fv{\mc X_n}\right|_{\tau_\P}
	=	\fv{X_n}
	=	\bmat{-\ab{\bs X_n}\\ \bs X_n},
		\label{Xn before}
}
due to the PLC condition $\lip{\fv{X_n}}^2=0$, derived from $\lip{\fv{X_n}-\fv{X_\P}}^2=0$ with $\fv X_\P=0$.\footnote{
One might think that the PLC condition $\lip{\fv{X_n}}^2=0$ leads to $\lip{\fv{X_n},\fv{U_n}}\stackrel{?}{=}0$ by differentiating it with respect to $\tau_n$.
This is not the case because the $n$th-particle's trajectory is not on the $\PLC\fn{x_\P}$ but away from it.
That is, $\lip{\fv{X_n}}^2=0$ does not hold along the path of the $n$th particle.
}
After the infinitesimal time evolution, the $n$th particle moves onto $\PLC\fn{\fv{x_\P'}}$.
In the $\mc C\fn{\tau_\P}$ frame,
\al{
\fv{X_n}=\bmat{-\ab{\bs X_n}\\ \bs X_n}
	&\longmapsto
		\fv{X_n'}=\fv{X_n}+\fv{U_n}\,\Delta\tau_n,\\
\fv{U_n}=\bmat{\gamma\fn{\bs U_n}\\ \bs U_n}
	&\longmapsto
		\fv{U_n'}=\fv{U_n}+\fv{A_n}\,\Delta\tau_n,
}
where we see from Eq.~\eqref{delta tau n limit} that\footnote{
As Eq.~\eqref{delta tau n limit} is manifestly Lorentz invariant, it is applicable in any frame.
}
\al{
\Delta\tau_n
	&\approx
		{\lip{\fv{X_n}-\fv{X_\P},\fv{U_\P}}\ov\lip{\fv{X_n}-\fv{X_\P},\fv{U_n}}}\,\Delta\tau_\P
	=	{\Delta\tau_\P\ov \gamma\fn{\bs U_n}+\bh X_n\cdot\bs U_n},
		\label{delta tau n from delta tau P}
}
in which we have used $\fv{X_\P}=\fv 0$, $\fv{U_\P}=\paren{1,\bs 0}$, and $\fv{X_n}=\paren{-\ab{\bs X_n},\bs X_n}=\ab{\bs X_n}\paren{-1,\bh X_n}$.
We see that a particle coming toowards the player, $\bh X_n\cdot\bs U_n<0$, spends a longer proper time compared to a particle going away from the player, $\bh X_n\cdot\bs U_n>0$; see Fig.~\ref{delta tau n figure}.

The Lorentz transformation from the $\mc C\fn{\tau_\P}$ frame to the $\mc C\fn{\tau_\P'}$ frame is
\al{
L\fn{\bs U_\P'}
	=	
L\fn{\bs A_\P\Delta\tau_\P}
	&=	\bmat{
			\gamma\fn{\bs A_\P\Delta\tau_\P}&-\bs A_\P^\T\Delta\tau_\P\\
			-\bs A_\P\Delta\tau_\P&\I+\paren{\gamma\fn{\bs A_\P\Delta\tau_\P}-1}\bh A_\P^\T\bh A_\P
			}\nn
	&\approx
		\bmat{
			1&-\bs A_\P^\T\Delta\tau_\P\\
			-\bs A_\P\Delta\tau_\P&\I
			},
}
and the co-moving coordinates $\fv{\mc X'}$ at $\tau_\P'$ are obtained from $\fv{X'}$ in the $\mc C\fn{\tau_\P}$ frame by the following Poincar\'e transformation:
\al{
\left.\fv{\mc X'}\right|_{\tau_\P'}
	&=	L\fn{\bs U_\P'}\paren{\fv{X'}-\fv{X_\P'}}	&
	&\approx
		\bmat{X^{\pr0}\\ \bs X'}
		-\bmat{1+\bs A_\P\cdot\bs X'\\ X^{\pr0}\bs A_\P}\Delta\tau_\P.
}
Accordingly, any velocity $\fv{U'}$ in the $\mc C\fn{\tau_\P}$ frame is transformed as
\al{
\left.\fv{\mc U'}\right|_{\tau_\P'}
	&=	L\fn{\bs U_\P'}\fv{U'}	&
	&\approx
		\bmat{U^{\pr0}\\ \bs U'}
		-\bmat{
			\bs A_\P\cdot\bs U'\\
			U^{\pr0}\bs A_\P
			}\Delta\tau_\P.
}
Putting $\fv{X_n'}$ and $\fv{U_n'}$ into $\fv{X'}$ and $\fv{U'}$ in the above expression, respectively, we get
\al{
\left.\fv{\mc X_n'}\right|_{\tau_\P'}
	&=	L\fn{\bs U_\P'}\paren{\fv{X_n'}-\fv{X_\P'}}	&
	&\approx
		\bmat{-\ab{\bs X_n}\\ \bs X_n}
		+\bmat{\gamma\fn{\bs U_n}\\ \bs U_n}\Delta\tau_n
		+\bmat{
			-\paren{1+\bs A_\P\cdot\bs X_n}\\
			\ab{\bs X_n}\bs A_\P }\Delta\tau_\P,
			\label{Xn after}
		\\
\left.\fv{\mc U_n'}\right|_{\tau_\P'}
	&=	L\fn{\bs U_\P'}\fv{U_n'}&
	&\approx
		\bmat{U_n^0\\ \bs U_n}
		+
		\bmat{A_n^0\\ \bs A_n}\Delta\tau_n
		-
		\bmat{
			\bs A_\P\cdot\bs U_n\\
			\gamma\fn{\bs U_n}\bs A_\P
			}\Delta\tau_\P,
}
where $\Delta\tau_n$ is given in Eq.~\eqref{delta tau n from delta tau P}.

To summarize, the basic evolution equation in the co-moving frames is
\al{
{\df\bs{\mc X}_n\ov\df\tau_\P}
	&=	{\bs{\mc U}_n\ov\gamma\fn{\bs{\mc U}_n}+\bh{\mc X}_n\cdot\bs{\mc U}_n}
		+\ab{\bs{\mc X}_n}\bs{\mc A}_\P,
			\label{co-moving X}\\
{\df\bs{\mc U}_n\ov\df\tau_\P}
	&=	{\bs{\mc A}_n\ov\gamma\fn{\bs{\mc U}_n}+\bh{\mc X}_n\cdot\bs{\mc U}_n}
		-\gamma\fn{\bs{\mc U}_n}\bs{\mc A}_\P,
			\label{co-moving U}
}
where
\al{
\left.{\df\bs{\mc X}_n\ov\df\tau_\P}\right|_{\tau_\P}
	&:=	\lim_{\Delta\tau_\P\to0}{\left.\bs{\mc X}_n'\right|_{\tau_\P'}-\left.\bs{\mc X}_n\right|_{\tau_\P}\ov\Delta\tau_\P}, 
}
with $\left.\bs{\mc X}_n'\right|_{\tau_\P'}$ and $\left.\bs{\mc X}_n\right|_{\tau_\P}$ being given in Eqs.~\eqref{Xn after} and \eqref{Xn before}, respectively, and similarly for $\bs{\mc U}_n$; we have used $\left.\bs{\mc U}_n\right|_{\tau_\P}=\bs U_n$ and $\left.\bs{\mc A}_n\right|_{\tau_\P}=\bs A_n$; and we have abbreviated the symbol $\left.\mbox{}\right|_{\tau_\P}$ in the left- and right-hand sides of Eqs.~\eqref{co-moving X} and \eqref{co-moving U}.

The second terms in the right-hand sides of Eqs.~\eqref{co-moving X} and \eqref{co-moving U} are the extra contribution from the Lorentz transformation that pulls back to the player's central frame in every time step.
When the player accelerates towards $\bs{\mc X}_n$, the $n$th-particle's apparent velocity~\eqref{co-moving X} receives the extra contribution $\ab{\bs{\mc X}_n}\bs{\mc A}_\P$ that kicks the $n$th particle away from the player.
If you accelerate towards an object, its apparent position becomes farther away from you!
On the other hand, the Lorentz transformation affects the apparent velocity~\eqref{co-moving U} by $-\gamma\fn{\bs{\mc U}_n}\bs{\mc A}_\P$, so the $n$th particle's apparent velocity tends to be pulled back to the player.
These two effects compete with each other, and should eventually bring the $n$th particle back to the player if the player continues to accelerate towards it.

Let us see this in the following way.
For simplicity, we let the $n$th particle stay at rest in the reference frame: $\bs{\mc A}_n=\bs 0$; see footnote~\ref{acceleration footnote}. We consider a constant acceleration of the player towards the $n$th particle in the player's rest frame:
\al{
\bs{\mc A}_\P
	&=	a\bh{\mc X}_n,
}
where $a>0$ is a constant. 
Note that $\bh{\mc A}_\P=\bh{\mc X}_\P$.
Putting this into Eq.~\eqref{co-moving U}, we obtain
\al{
{\df\bs{\mc U}_n\ov\df\tau_\P}
	&=	-\gamma\fn{\bs{\mc U}_n}a\bh{\mc X}_n
		\label{dUdtau}
}
We set the initial condition such that the player had also been at rest in the reference frame before its acceleration.
Then the direction $\bh{\mc X}_n$ remains constant throughout the time evolution until the $n$th particle collides with the player, and the apparent velocity of the $n$th particle is proportional to its direction: $\bh{\mc U}_n\propto\bh{\mc X}_n$.
We parametrize the apparent position and velocity as
\al{
\bs{\mc X}_n\fn{\tau_\P}
	&=	\mc X\fn{\tau_\P}\bh{\mc X}_n,\\
\bs{\mc U}_n\fn{\tau_\P}
	&=	\mc U\fn{\tau_\P}\bh{\mc X}_n,
}
where $\mc U\fn{\tau_\tx{P}}$ is positive and negative when the $n$th particle appears to be moving away and coming back to the player, respectively. Then Eq.~\eqref{dUdtau} reads
\al{
{\df\mc U\ov\df\tau_\P}
	&=	-a\sqrt{1+\mc U^2},
}
which has the solution
\al{
\mc U\fn{\tau_\P}
	&=	-\sinh\fn{a\tau_\tx{P}}
}
under the initial condition $\mc U\fn{0}=0$.
Putting this into Eq.~\eqref{co-moving X}, we obtain
\al{
{\df\mc X\ov\df\tau_\P}
	&=	-e^{a\tau_\P}\sinh\fn{a\tau_\P}+a\mc X,
}
which is solved as
\al{
\mc X\fn{\tau_\P}
	&=	e^{a\tau_\P}\paren{\mc X_0+{1-\cosh\fn{a\tau_\P}\ov a}},
		\label{apparent distance in NU}
}
under the initial condition $\mc X\fn{0}=\mc X_0$, where $\mc X_0>0$ is the initial apparent distance.
The apparent velocity and acceleration read
\al{
{\df\mc X\fn{\tau_\P}\ov\df\tau_\P}
	&=	e^{a\tau_\P}\paren{1+a\mc X_0-e^{a\tau_\P}},\\
{\df^2\mc X\fn{\tau_\P}\ov{\df\tau_\P}^2}
	&=	a\,e^{a\tau_\P}\paren{1+a\mc X_0-2e^{a\tau_\P}}.
		\label{apparent acceleration in NU}
}

We see that the particle indeed appears to be moving away from the player at the beginning: The apparent velocity is positive while $\tau_\P<\tau_\P^c$, with the critical value $\tau_\P^c$ being given by
\al{
\tau_\P^c
	&:=	{\ln\fn{1+a\mc X_0}\ov a}
	=	{c\ov a}\ln\fn{1+{a\mc X_0\ov c^2}},
}
where, in the last step, we have recovered the speed of light from natural units.
The apparent distance at $\tau_\P^c$ is, in nonnatural units,
\al{
\mc X\fn{\tau_\P^c}
	&=	\mc X_0+{a\mc X_0^2\ov 2c^2}.
}

When the apparent distance is small, $\mc X_0\ll c^2/a$, we get $\tau_\P^c\to\mc X_0/c$.
Then if one wants to see this effect for one second, the apparent distance must be larger than $3\times 10^8\,\tx{m}$, which is roughly the distance between the Earth and Moon $\simeq 4\times 10^8\,\tx{m}$. Even when the apparent distance is of that order, an acceleration of the order of the gravitational acceleration on Earth, $a\sim10\,\tx{m}/\tx{s}^2$, elongates $\mc X\fn{\tau_\P^c}$ from $\mc X_0$ only by a fraction of $a\mc X_0/2c^2\sim 10^{-8}$, i.e., by $a\mc X_0^2/2c^2\sim 10\,\tx{m}$. This is a tiny effect for the distance and acceleration of everyday life. In computer games, one may consider a very large distance and/or acceleration to see such a relativistic effect.

The final proper time of collision is
\al{
\tau_\P^f
	&=	{c\ov a}\arccosh\fn{1+{a\mc X_0\ov c^2}}
	=	\begin{cases}
		{c\ov a}\ln\fn{2a\mc X_0\ov c^2}+\Or{\paren{a\mc X_0\ov c^2}^{-1}}
			&	\tx{for $a\mc X_0\gg c^2$,}\\
		\sqrt{2\mc X_0\ov a}+\Or{\paren{a\mc X_0\ov c^2}^{3/2}}
			&	\tx{for $a\mc X_0\ll c^2$.}
		\end{cases}
}
That is, the proper-time difference between the turnover and collision is
\al{
\tau_\P^f-\tau_\P^c
	&=	\begin{cases}
		{c\ov a}\ln2+\Or{\paren{a\mc X_0\ov c^2}^{-2}}
			&	\tx{for $a\mc X_0\gg c^2$,}\\
		\sqrt{\mc X_0\ov a}+\Or{a\mc X_0\ov c^2}
			&	\tx{for $a\mc X_0\ll c^2$.}
		\end{cases}
}
The corresponding apparent speed is
\al{
{\mc X\fn{\tau_\P^c}\ov\tau_\P^f-\tau_\P^c}
	&=	\begin{cases}
		{c\ov 2\ln2}\paren{a\mc X_0\ov c^2}^2+\Or{a\mc X_0\ov c^2}
			&	\tx{for $a\mc X_0\gg c^2$,}\\
		c\sqrt{a\mc X_0\ov 2c^2}+\Or{a\mc X_0\ov c^2}
			&	\tx{for $a\mc X_0\ll c^2$,}
		\end{cases}
}
which exceeds the speed of light when $a\mc X_0\gtrsim c^2$.
Note that this does not lead to any contradiction: The distance for a given proper-time interval can be arbitrarily large, as is the case for the covariant velocity discussed in Sec.~\ref{proper time and covariant velocity}.

Recovering the speed of light $c$, we may rewrite Eqs.~\eqref{apparent distance in NU}--\eqref{apparent acceleration in NU} in nonnatural units:
\al{
\mc X\fn{\tau_\P}
	&=	e^{a\tau_\P/c}\paren{\mc X_0+{c^2\ov a}\sqbr{1-\cosh\fn{a\tau_\P\ov c}}},
		\label{apparent distance in NN}\\
{\df\mc X\fn{\tau_\P}\ov\df\tau_\P}
	&=	c\,e^{a\tau_\P/c}\paren{1+{a\mc X_0\ov c^2}-e^{a\tau_\P/c}},
		\label{apparent velocity}\\
{\df^2\mc X\fn{\tau_\P}\ov{\df\tau_\P}^2}
	&=	a\,e^{a\tau_\P/c}\paren{1+{a\mc X_0\ov c^2}-2e^{a\tau_\P/c}}.
		\label{apparent acceleration in NN}
}
In a formal expansion of Eqs.~\eqref{apparent distance in NN}--\eqref{apparent acceleration in NN}, assuming that $c=1$ were large, we obtain
\al{
\mc X\fn{\tau_\P}
	&=	\mc X_0-{a\ov2}\tau_\P^2+\Or{c^{-1}},\\
{\df\mc X\fn{\tau_\P}\ov\df\tau_\P}
	&=	-a\tau_\P+\Or{c^{-1}},\\
{\df^2\mc X\fn{\tau_\P}\ov{\df\tau_\P}^2}
	&=	-a+\Or{c^{-1}},
}
which coincides with the result for the constant acceleration in ordinary Newtonian mechanics.
On the other hand, the correct limit for $c\gg a\tau_\P$ reads
\al{
\mc X\fn{\tau_\P}
	&=	\mc X_0+c\tau_\P{a\mc X_0\ov c^2}
		-{a\ov2}\tau_\P^2\paren{1-{a\mc X_0\ov c^2}}
		+\Or{\paren{a\tau_\P\ov c}^3},\\
{\df\mc X\fn{\tau_\P}\ov\df\tau_\P}
	&=	c{a\mc X_0\ov c^2}-a\tau_\P\paren{1-{a\mc X_0\ov c^2}}+\Or{\paren{a\tau_\P\ov c}^2},\\
{\df^2\mc X\fn{\tau_\P}\ov{\df\tau_\P}^2}
	&=	-a\paren{1-{a\mc X_0\ov c^2}}+\Or{a\tau_\P\ov c}.
}
Again we see that the relativistic correction is significant for $\mc X_0\gtrsim c^2/a$, as well as for $\mc X_0\gtrsim c\tau_\P$.

\section{Rocket propulsion}\label{rocket propulsion}
As a concrete example of the relativistic acceleration, we show how the rocket propulsion is formulated.

Suppose the following:
A rocket has a mass $m\fn{\tau}$, is located at $\fv x\fn{\tau}$, and has a velocity $\fv u\fn{\tau}$ in a reference frame $\Set{\fv x}$ at its proper time $\tau$.
Note that a mass is a Lorentz-invariant notion and does not depend on frames.
The rocket is in an inertial frame, i.e., under no outer force.

Let us first switch to the rest frame of the rocket at $\tau$:
\al{
\fv X
	&=	L\fn{\bs u\fn{\tau}}\fv x,\\
\fv U\fn{\tau}
	&=	L\fn{\bs u\fn{\tau}}\fv u\fn{\tau}
	=	\bmat{1\\ \bs 0}.
		\label{U in rest frame}
}
Suppose that the rocket's mass, including its fuel, has changed by an amount
\al{
\Delta m<0
}
during an infinitesimal proper time $\Delta\tau$.
That is, its fuel is consumed by an amount $\ab{\Delta m}$.
At the proper time $\tau+\Delta\tau$
the rocket's mass becomes $m+\Delta m$, with $\Delta m<0$, and 
its velocity
\al{
\fv U\fn{\tau+\Delta\tau}=\fv U\fn{\tau}+\fv{\Delta U}.
}
Note that Eq.~\eqref{u normalization} implies that $\lip{\fv U+\fv{\Delta U}}^2=-1$, which leads to\footnote{
We may instead directly take the derivative of $\lip{U}^2=-1$ to obtain $\lip{{\df U\over\df\tau},U}=0$, which is also true in any frame; see Eq.~\eqref{u dot a}.
}
\al{
\lip{\fv U\fn{\tau},\,\fv{\Delta U}}
	&=	0.
		\label{U orthogonality}
}
This relation applies in any frame.
In the rest frame~\eqref{U in rest frame}, we further obtain
\al{
\Delta U^0
	&=	0.
		\label{deltaU0}
}

Suppose that the fuel $\ab{\Delta m}$ is converted into a jet of mass $\Delta m_\tx{jet}>0$ with velocity $\fv W$.
The energy and $d$-momentum conservation reads
\al{
mU^0
	&=	\paren{m+\Delta m}\paren{U^0+\Delta U^0}+\Delta m_\tx{jet}W^0+\Delta E_\tx{dis},\\
m\bs U
	&=	\paren{m+\Delta m}\paren{\bs U+\Delta \bs U}+\Delta m_\tx{jet}\bs W,
}
where $\Delta E_\tx{dis}>0$ is a possible energy dissipation.\footnote{
For example, $\Delta E_\tx{dis}$ includes the part of kinetic energy of the jet constituents that comes from the velocity components that are perpendicular to the (much larger) net jet velocity and cancel each other as a whole. 
}
Putting $U^0=1$, $\bs U=0$, $\Delta U^0=0$, and $W^0=\sqrt{1+\bs W^2}$ and dropping the quadratic infinitesimal terms, we obtain
\al{
\ab{\bs W}
	&=	\sqrt{\paren{1-C_\tx{dis}\ov C_\tx{jet}}^2-1},\\
\Delta\bs U
	&=	{\Delta m\ov m}\sqrt{\paren{1-C_\tx{dis}^2}-C_\tx{jet}^2}\,\bh W,
}
where we have assumed that $C_\tx{jet}:=-\Delta m_\tx{jet}/\Delta m>0$ and $C_\tx{dis}:=-\Delta E_\tx{dis}/\Delta m>0$ are constants.

The final expression for the rocket's $D$ acceleration in its rest frame is
\al{
{\df\fv U\fn{\tau}\over\df\tau}
	&=	\sqrt{\paren{1-C_\tx{dis}}^2-C_\tx{jet}^2}\,
		{\df \ln m\fn{\tau}\over\df\tau}\bmat{
			0\\
			\bh W\fn{\tau}
			},
		\label{rocket acceleration in rest frame}
}
where ${\df \ln m\ov\df\tau}={1\ov m}{\df m\ov\df\tau}$.
Since Eq.~\eqref{rocket acceleration in rest frame} is written Lorentz covariantly, we can easily obtain its form in the reference frame at $\tau$:
\al{
{\df\fv u\fn{\tau}\over\df\tau}
	&=	\sqrt{\paren{1-C_\tx{dis}}^2-C_\tx{jet}^2}\,
		{\df \ln m\fn{\tau}\over\df\tau}\,
		L\fn{-\bs u\fn{\tau}}
		\bmat{
			0\\
			\bh W\fn{\tau}
			}.
		\label{rocket acceleration}
}
To summarize, when one switches on the rocket and determines its thrust direction to be $\bh W$ in its rest frame, the given acceleration is Eq.~\eqref{rocket acceleration} in the general reference frame.
In terms of the acceleration in Eq.~\eqref{self acceleration}, the rocket propulsion corresponds to setting
\al{
A_s=\sqrt{\paren{1-C_\tx{dis}}^2-C_\tx{jet}^2}\,
		{\df \ln m\fn{\tau}\ov\df\tau}.
}

The (noncovariant) exhaust velocity of the jet is
\al{
V	&=	{\ab{\bs W}\ov\sqrt{1+\bs W^2}}
	=	\sqrt{1-\paren{C_\tx{jet}\ov1-C_\tx{dis}}^2}.
		\label{exhaust velocity}
}
When the jet is composed of massless photons, we may take the limit $\Delta m_\tx{jet}\to0$ i.e.\ $C_\tx{jet}\to0$, in which the exhaust velocity approaches the speed of light $V\to1$.
In the limit, Eq.~\eqref{rocket acceleration} reads
\al{
\left.{\df\fv u\fn{\tau}\over\df\tau}\right|_{C_\tx{jet}\to0}
	&=	\paren{1-C_\tx{dis}}
		{\df \ln m\fn{\tau}\over\df\tau}\,
		L\fn{-\bs u\fn{\tau}}
		\bmat{
			0\\
			\bh W\fn{\tau}
			}.
}

As an illustration, let us examine a rocket propulsion with the constant direction $\bh W\fn{\tau}=\paren{0,0,-1}$.
We set an initial condition $\bs u\fn{0}=\bs 0$ and $m\fn{0}=m_\tx{ini}>0$.
For a velocity $\fv u=\paren{\sqrt{1+u^2},0,0,u}$, the Lorentz transformation to the rest frame~\eqref{to rest frame} is\footnote{
Here and hereafter in this section, $u^2$ is not the second component of $\fv u$ but the square of the parameter~$u$.
}
\al{
L\fn{\bs u}
	&=	\bmat{
			\sqrt{1+u^2}&0&0&-u\\
			0&1&0&0\\
			0&0&1&0\\
			-u&0&0&\sqrt{1+u^2}
			},
			\label{z-boost}
}
and the $D$ acceleration is
\al{
{\df\fv u\fn{\tau}\over\df\tau}
	&=	\sqrt{\paren{1-C_\tx{dis}}^2-C_\tx{jet}^2}\,
		{\df \ln m\fn{\tau}\over\df\tau}
		\bmat{-u\\
				0\\0\\
				-\sqrt{1+u^2}}.
}
Solving the spatial component,
\al{
\int_0^{u_\tx{fin}}{\df u\over \sqrt{1+u^2}}
	&=	-\sqrt{\paren{1-C_\tx{dis}}^2-C_\tx{jet}^2}
		\int_{m_\tx{ini}}^{m_\tx{fin}}\df\ln m,
}
we obtain the relation between the final velocity $u_\tx{fin}$ and mass $m_\tx{fin}$:
\al{
u_\tx{fin}
	&=	\sinh\fn{
			\sqrt{\paren{1-C_\tx{dis}}^2-C_\tx{jet}^2}\,
			\ln{m_\tx{ini}\over m_\tx{fin}}}.
		\label{u vs m}
}
We see that the ratio of the initial rocket mass, $m_\tx{ini}$, to the final mass after using up all the fuel, $m_\tx{fin}$, governs the final covariant velocity $u_\tx{fin}$.
When we set $C_\tx{dis}=0$, the exhaust velocity~\eqref{exhaust velocity} becomes $V=\sqrt{1-C_\tx{jet}^2}$, and the small velocity limit $V\ll1$ of Eq.~\eqref{u vs m} recovers the well known result $u_\tx{fin}\to V\ln{m_\tx{ini}\ov m_\tx{fin}}$.

In the limit $m_\tx{fin}\ll m_\tx{ini}$, we can increase the final covariant velocity $u_\tx{fin}\to{1\ov2}\paren{m_\tx{ini}/ m_\tx{fin}}^{\sqrt{\paren{1-C_\tx{dis}}^2-C_\tx{jet}^2}}$ without limit, and the rocket's noncovariant velocity approaches the speed of light $v_\tx{fin}=\tanh\fn{\sqrt{\paren{1-C_\tx{dis}}^2-C_\tx{jet}^2}\,\ln{m_\tx{ini}\over m_\tx{fin}}}\to 1$.

\section{Handling rotation with quaternions}\label{quaternion}
We review how to implement a spatial rotation in three dimensions using the quaternion, which avoids the gimbal lock. This can be used to store the information on the direction of each 3D object introduced in Sec.~\ref{rigid body section}.

Let $\bs q$ be a three-vector:
\al{
\bs q=\paren{q^1,q^2,q^3}=\bmat{q^1\\ q^2\\ q^3}.
}
We parametrize a quaternion $Q$ by four real numbers $q$, $q^1$, $q^2$ and $q^3$ as
\al{
Q	&=	\paren{q;\,q^1,\,q^2,\,q^3}
	=	q+q^1\iq_1+q^2\iq_2+q^3\iq_3
	=:	q+\bs q\cdot\bs\iq,
}
where the unit quaternions obey
\al{
\iq_1^2
	=	\iq_2^2=\iq_3^2=\iq_1\iq_2\iq_3
	=	-1,
}
in which $\iq_i^2:=\paren{\iq_i}^2=\iq_i\iq_i$.
That is,
\al{
\iq_i\iq_j
	&=	-\delta_{ij}+\sum_{k=1}^3\epsilon_{ijk}\iq_k,
		\label{basis relation}
}
where
$\epsilon_{ijk}$ is the totally antisymmetric tensor\footnote{
To be explicit, the nonzero components of the totally antisymmetric tensor are $\epsilon_{123}=\epsilon_{231}=\epsilon_{312}=1$ and $\epsilon_{132}=\epsilon_{321}=\epsilon_{213}=-1$. All the other components are zero.
}
\al{
\epsilon_{ijk}
	&=	\begin{cases}
		1	&	\tx{for $\paren{i,j,k}$ being even permutation of $\paren{1,2,3}$,}\\
		-1	&	\tx{for $\paren{i,j,k}$ being odd permutation of $\paren{1,2,3}$,}\\
		0	&	\tx{otherwise.}
		\end{cases}
}
Hereafter, we employ the shorthand notation $\sum_{ij}=\sum_{i=1}^3\sum_{j=1}^3$ etc.
From the relation~\eqref{basis relation}, it is straightforward to show that
\al{
PQ	=	\paren{p+\sum_ip^i\iq_i}\paren{q+\sum_jq^j\iq_j}
	&=	\paren{pq-\sum_ip^iq^i}
		+\sum_k\paren{qp^k+pq^k+\sum_{ij}\epsilon_{ijk}p^iq^j}\iq_k\nn
	&=	\paren{pq-\bs p\cdot\bs q}
		+\paren{q\bs p+p\bs q+\bs p\times\bs q}\cdot \bs\iq,
}
where the vector product is defined by $\paren{\bs p\times\bs q}_i=\sum_{jk}\epsilon_{ijk}p^jq^k$.
In particular,
\al{
\paren{\bs p\cdot\bs\iq}^2
	&=	-\bs p^2.
}
For any $Q=\paren{q;q^1,q^2,q^3}$, we define its conjugate $\ol Q$ by
\al{
\ol Q
	&=	\paren{q;-q^1,-q^2,-q^3}
	=	q-\sum_iq^i\iq_i
	=	q-\bs q\cdot\bs\iq.
}
Note that
\al{
\ol{PQ}
	&=	\ol Q\,\ol P.
}

We express a three-vector, say, the Cartesian coordinates $\bs x=\paren{x^1,x^2,x^3}$ by the quaternion\footnote{
Here and hereafter in this section, $X$ stands for the quaternion~\eqref{quaternion X} rather than the rest frame coordinate.
}
\al{
X	&=	\paren{0;x^1,x^2,x^3}
	=	\sum_ix^i\iq_i=x^1\iq_1+x^2\iq_2+x^3\iq_3.
		\label{quaternion X}
}
We want to rotate this vector by an angle $\theta$ around an axis $\bs n=\paren{n^1,n^2,n^3}$. 
The resultant three-vector can be obtained from
\al{
X'	&=	NX\ol N,
}
where
\al{
N
	&:=	\cos{\theta\over2}+\paren{\sin{\theta\over 2}}{\bs n\cdot\bs\iq}.
		\label{rotation quaternion}
}
One may use the quaternion~\eqref{rotation quaternion} to specify the direction of a rigid body.

It is tedious but straightforward to show that\footnote{
One uses the identity
\als{
\sum_i\epsilon_{ijk}\epsilon_{ij'k'}=\delta_{jj'}\delta_{kk'}-\delta_{jk'}\delta_{kj'}
}
copiously.
}
\al{
NX\ol N
	&=	\sqbr{
			\paren{c^2-s^2\bs n^2}\bs x
			+2sc\,\bs n\times\bs x
			+2s^2\paren{\bs n\cdot\bs x}\bs n
			}\cdot\bs\iq,
}
where $s:=\sin{\theta\over2}$ and $c:=\cos{\theta\over2}$.
When we take $\bs n$ to be a unit vector $\bh n$ with $\ab{\bh n}=1$, we obtain further
\al{
NX\ol N
	&=	\sqbr{
			\paren{\cos\theta}\bs x
			+\paren{\sin\theta}\bh n\times\bs x
			+\paren{1-\cos\theta}\paren{\bh n\cdot\bs x}\bh n
			}\cdot\bs\iq.
}
Suppose that we decompose $\bs x$ into the components that are parallel and perpendicular to the unit vector $\bh n$:
\al{
\bs x
	&=	\bs x_\parallel+\bs x_\perp,
}
where
\al{
\bs x_\parallel
	&:=	\paren{\bh n\cdot\bs x}\bh n,	&
\bs x_\perp
	&:=	-\bh n\times\paren{\bh n\times\bs x}.
}
Then\footnote{
The orientation of $\bh x_\parallel$, $\bh x_\perp$ and $\bh n\times\bh x$, which are perpendicular to each other, is such that
\als{
\bh x_\perp\times\paren{\bh n\times\bh x}
	&=	\bh x_\parallel
	=	\bh n,
}
where $\bh x:=\bs x/\ab{\bs x}$.
}
\al{
NX\ol N
	&=	\sqbr{
			\bs x_\parallel
			+\paren{\cos\theta}\bs x_\perp
			+\paren{\sin\theta}\bh n\times\bs x
			}\cdot\bs\iq.
}
In other words, the three-vector $\bs x$ is transformed by $X\to X'=NX\ol N$ as
\al{
x^i	&\to	x^{\prime i}=\sum_jR^{ij}x^j,
}
where
\al{
R^{ij}
	&=	\h n_i\h n_j\paren{1-\cos\theta}
		+\delta_{ij}\cos\theta
		-\sum_k\epsilon_{ijk}\h n_k\sin\theta.
}

\bibliography{relativity}
\bibliographystyle{TitleAndArxiv}
\end{document}